\documentclass{article}

\usepackage[preprint,nonatbib]{neurips_2024}

\usepackage[utf8]{inputenc} 
\usepackage[T1]{fontenc}    
\usepackage{hyperref}       
\usepackage{url}            
\usepackage{booktabs}       
\usepackage{amsfonts}       
\usepackage{nicefrac}       
\usepackage{microtype}      
\usepackage{xcolor}         
\usepackage{amsthm}

\newtheorem{theorem}{Theorem}

\usepackage{graphicx}
\usepackage[style=ieee, citestyle=numeric-comp, backend=biber]{biblatex}
\usepackage{amsmath}
\usepackage{amssymb}
\usepackage{bbm}
\usepackage[toc,page]{appendix}
\usepackage{wrapfig}
\usepackage{floatrow}

\newcommand{\bmm}[1]{\ensuremath{\mathbf{#1}}}

\usepackage[mathscr]{eucal}
\usepackage{authblk}
   
\title{Binding in hippocampal-entorhinal circuits \\ enables compositionality in cognitive maps}

\def\R{{\mathbb{R}}}
\def\pr{{\rm Pr}}
\def\E{{\mathbb E}}

\def\C{{\mathbb C}}

\author[1*]{Christopher J. Kymn}
\author[1,2]{Sonia Mazelet}
\author[1]{Anthony Thomas}
\author[3]{Denis Kleyko}
\author[4]{E. Paxon Frady}
\author[1,4]{Friedrich T. Sommer}
\author[1]{Bruno A. Olshausen}

\affil[1]{Redwood Center for Theoretical Neuroscience, UC Berkeley, Berkeley, USA}
\affil[2]{Université Paris-Saclay, ENS Paris-Saclay, Gif-sur-Yvette, France}
\affil[3]{Centre for Applied Autonomous Sensor Systems, Örebro University, Örebro, Sweden}
\affil[4]{Intel Labs, Santa Clara, USA}
\affil[*]{cjkymn@berkeley.edu}

\begin{document}

\maketitle

\begin{abstract}
We propose a normative model for spatial representation in the hippocampal formation that combines optimality principles, such as maximizing coding range and spatial information per neuron, with an algebraic framework for computing in distributed representation. Spatial position is encoded in a residue number system, with individual residues represented by high-dimensional, complex-valued vectors. These are composed into a single vector representing position by a similarity-preserving, conjunctive vector-binding operation. Self-consistency between the representations of the overall position and of the individual residues is enforced by a modular attractor network whose modules correspond to the grid cell modules in entorhinal cortex. The vector binding operation can also associate different contexts to spatial representations, yielding a model for entorhinal cortex and hippocampus. We show that the model achieves normative desiderata including superlinear scaling of patterns with dimension, robust error correction, and hexagonal, carry-free encoding of spatial position. These properties in turn enable robust path integration and association with sensory inputs.
More generally, the model formalizes how compositional computations could occur in the hippocampal formation and leads to testable experimental predictions.
\end{abstract}

\section{Introduction}
The hippocampal formation (HF), consisting of hippocampus (HC) and the medial and lateral part of the neighboring entorhinal cortex, (MEC) and (LEC), is critical for forming memories and representing variables such as spatial position~\cite{eichenbaum2017integration,moser2017spatial}. Recent work has provided evidence of compositional structure in HF representations, for example, novel recombinations of past experience occurring in replay~\cite{kurth2023replay}, or the exponential expressivity of the grid cell code~\cite{fiete2008grid, sreenivasan2011grid}. In particular, compositional representations afford high expressivity with lower dimensional storage requirements~\cite{behrens2018cognitive}, less complexity in latent state inference, and generalization to novel scenes with familiar parts. 

To gain insight into the possible computational principles and neural mechanisms at play in the HF, we take a normative modeling approach. That is, we seek a set of neural coding principles that effectively achieve the postulated function of the system. With this approach, we can then explain details about the neuroanatomical and neurophysiological structures in light of their particular contributions to an information processing objective. We believe that the resulting model can also lead to new predictions about the neural mechanisms that enable this function.

The postulated function of the HF ---as a cognitive map and episodic memory--- has a core computational requirement, to represent and navigate space. Here, space is either the actual physical environment or a more abstract conceptual space. We formulate multiple desiderata for an effective representation of space. We then show that a residue number system, incorporated into a compositional encoding scheme, fulfills these desiderata. It is achieved by a modular attractor network that factorizes the individual components of encoded locations. This provides an algorithmic-level hypothesis of hippocampal-entorhinal interactions. A core mechanism of this algorithm is \textit{binding}, which draws inspiration from work in neuroscience, cognitive science, and artificial intelligence.

\section{A normative model for the hippocampal formation}

\vspace{-0.05in}
\subsection{Principles for representing space}
\label{sec:space_representation}

Our first set of normative requirements is that space is represented by a compositional code that has high spatial resolution, is noise-robust, and in which algebraic operations on the components can be updated in parallel. 
Prior work~\cite{fiete2008grid,sreenivasan2011grid} has proposed the residue number system (RNS) \cite{garner1959residue} as a candidate for fulfilling these requirements. An RNS expresses an integer $x$ in terms of its remainder relative to a set of co-prime moduli $\{m_i\}$. For example, relative to moduli $\{3, 5, 7\}$, $x=40$ is encoded as $\{1, 0, 5\}$. The Chinese Remainder Theorem guarantees that all integers in the range $[0,M-1]$, where $M = \prod_i m_i$, are assigned a unique representation.  
An RNS provides high spatial resolution, carry-free arithmetic operations, and robust error correction~\cite{goldreich1999chinese}.
Experimental observations in entorhinal cortex show a discrete multi-scale organization of spatial grid cells~\cite{stensola2012entorhinal} that is compatible with an organization into discrete RNS modules.

The second normative principle we adopt is that an individual residue value should be encoded by a neural population in a similarity-preserving fashion. In particular, we require that distinct integer values are represented with nearly orthogonal vectors.
To achieve this principle, we use a method similar to random Fourier features \cite{rahimi2007random}. Each modulus, with value $m_i$, is assigned a seed phasor vector, $\mathbf{g}_i \in \mathbb{C}^D$, whose elements $(\mathbf{g}_i)_j$ are drawn uniformly from the $m_i$-th roots of unity (i.e., $(\mathbf{g}_i)_j=e^{\sqrt{-1}\,\omega_{ij}}$, with $\omega_{ij}=\frac{2\pi}{m_i}\,k_j$, and $k_j$ chosen randomly from $\{0,...,m_i-1\}$). The representation of a particular residue value $a_i \in \{0,\dots,m_i-1\}$ is then given by rotating the phases of the seed vector according to~\cite{plate1992holographic}: 
\begin{equation}
\mathbf{g}_i(a_i) = (\mathbf{g}_i)^{a_i}, 
\label{eq:fpe}
\end{equation}
where we abuse notation slightly to also think of $\bmm{g}_{i}$ as a function that takes $a_{i}$ as input and produces an embedding as described above. The complex-valued vectors can be mapped to interpretable population vectors via a randomized Fourier transform (Figures \ref{fig:path_integration}D and \ref{fig:remapping}). 

Our third normative principle concerns the manner in which a unique representation of a particular point in space is formed from the individual residue representations.
This requires that we somehow combine the residue vectors for each modulus.
Combining via concatenation, though straightforward, is not effective because codes that coincide in subsets of their residue representation would be similar, even when the encoded values are very different. 
Thus, the method of combining residue codes must be {\em conjunctive}. 
Conjunctive composition is often called \emph{binding} and is of fundamental importance in neuroscience~\cite{MalsburgAssemblies1986}, cognitive science~\cite{FodorCritical1988}, and machine learning~\cite{greff2020binding}.
An early proposal for binding is the tensor product of representation vectors~\cite{SmolenskyTensor1990}, with the tensor order equal to the number of bound objects. 

Here, we implement binding with component-wise vector multiplication, a dimensionality preserving operation that represents a lossy compression of the full tensor product~\cite{plate1991holographic,kanerva2009hyperdimensional}. 
The resulting compositional vector representation of an integer $x \in \mathbb{Z}$ using an RNS representation with $K$ moduli, $\{a_1,a_2,..,a_K\}$, is: 
\vspace{-0.1in}
\begin{equation}
\mathbf{p}(x) = \bigodot_{i=1}^K \mathbf{g}_i(a_i).
\label{eq:rnsvecrep}
\end{equation}
We prove in Appendix~\ref{sec:residue_kernel} that this coding scheme represents distinct integer states using nearly orthogonal vectors, and that it generalizes in a natural way to support representation of arbitrary real numbers in a similarity preserving fashion.

Eq.~\ref{eq:rnsvecrep} represents individual points along a line. In general, however, a spatial representation involves points in 2D or 3D spaces. Conveniently, vector binding can be also used to compose representations of multidimensional lattices from vectors representing individual dimensions. As we will explain, there is still a choice in this composition that determines the resulting lattice structure. Following earlier proposals~\cite{wei2015principle,mathis2015probable,anselmi2020computational}, our fourth normative principle is to choose the lattice structure so that spatial information is maximized, as described in Section~\ref{sec:hex_main}.  

The final normative principle we require is that for computations such as path integration, there should be a simple vector manipulation that results in addition of the encoded variables. Again, vector binding provides this functionality with our coding strategy, because of the following property: 
\begin{equation}
\mathbf{g}(x) \odot \mathbf{g}(y) = \mathbf{g}(x+y).
\label{eq:varadd}
\end{equation}

\subsection{Modular attractor network for spatial representation}
\label{sec:modattspace}

A standard model of grid cell circuits is the line attractor, in which states that represent a consistent location lie on a low-energy manifold~\cite{fiete2008grid}. When initialized from a noisy location pattern, the circuit dynamics will generate a denoised location representation. Rather than forming a line attractor model for the entire representational space (Eq.~\ref{eq:rnsvecrep}), we propose a modular network architecture, so that the compositional structure of a residue number representation can scale towards a large range with fewer memory resources (Section~\ref{sec:empirical_scaling}), in a manner robust to noise (Section~\ref{sec:noise_robustness}). 

A starting point for our attractor network model is the Hopfield network, which acts as an associative memory by storing memory patterns as fixed-point attractors. The Rademacher-Hopfield network~\cite{hopfield1982neural} is a dynamical system whose state is a vector $\mathbf{x} \in \{-1,+1 \}^{D}$ that obeys the following dynamics:
\begin{equation}
    \mathbf{x}(t+1) = \text{sign} ( \mathbf{XX}^{T} \mathbf{x}(t) )
\end{equation}
with $\mathbf{X}$ as the matrix of memorized patterns (column vectors of $\mathbf{X}$).
The fixed-point attractor dynamics can be generalized to complex memory patterns $\mathbf{z} \in \mathbb{C}^{D}$:
\begin{equation}
    \mathbf{z}(t+1) = \sigma ( \mathbf{ZZ}^{\dag} \mathbf{z}(t) ),
    \label{eq:noest}
\end{equation}
where $\sigma$ is a non-linearity normalizing the amplitude of each complex-valued component to one~\cite{noest1987phasor}, and $\mathbf{Z}$ the corresponding matrix of memorized patterns. The model can also be discretized, such that each component is often quantized to a $r$-state phasor~\cite{noest1988discrete}. The Rademacher-Hopfield model is the special case where $r=2$ and the phasors happen to be real-valued.

An $r$-state phasor network of the form of Eq.~\ref{eq:noest} is well-suited to serve as an attractor network for each of the residue vectors in an RNS representation of position, with $r=m_i$ for modulus $i$, and the matrix $\mathbf{Z}$ (which we shall denote $\mathbf{G}_i$) storing the $\mathbf{g}_i(a_i)$ for $a_i\in\{0,..,m_i-1\}$. 
However, we desire a method for representing the whole coding range $M:= \prod_i^K m_i$ without storing all $M$ patterns in one large associative memory. For this purpose we show that a {\em resonator network}, a recently proposed recurrent network for \textit{unbinding} conjunctive codes~\cite{frady2020resonator,kent2020resonator}, lets us represent this range by storing only $n:=\sum_i^K m_i  \ll M$ patterns. 
Given a vector encoding of position, $\mathbf{p}(x)$, as formulated in Eq.~(\ref{eq:rnsvecrep}), a resonator network will factorize it into its constituent RNS components by iteratively updating each residue vector estimate, $\hat{\mathbf{g}}_i$, similar to the attractor dynamics of Eq.~(\ref{eq:noest}) but in a way that it is also consistent with $\mathbf{p}(x)$ given all other residue estimates $\hat{\mathbf{g}}_{j\neq i}$:
 \begin{equation}
    \mathbf{\hat{g}}_i(t+1) = \sigma \Bigl( \mathbf{G}_i \bmm{G}_i^{\dag} \bigl( \mathbf{p} \bigodot_{j \neq i}^K \mathbf{\hat{g}}_j^{*}(t) \bigr) \Bigr) \;\;\;\forall \ i
    \label{eq:reso}
\end{equation}

Let us now assume that the input $\mathbf{p}(x_t)$ encodes a spatial position $x_t$ using Eq.~(\ref{eq:rnsvecrep}). Given a velocity input $\mathbf{q}_i(v_t)$, estimated from self-motion input, path integration is performed by first running attractor dynamics, \textit{then} updating attractor states by velocity.
\begin{equation}
    \label{eqn:update-step}
    \mathbf{\hat{g}}_i(t+1) = \mathbf{q}_i(v_t) \odot \sigma ( \mathbf{G}_i \mathbf{G}_i^{\dag} \mathbf{p}(x_t) \bigodot_{i \neq j}^K \mathbf{\hat{g}}_j^{*}(t))
\end{equation}
After velocity updates, one can update the input state $\mathbf{p}(x_t)$ with the conjunctive representation of the current factor estimates: 
\begin{equation}
    \label{eqn:place-update}
    \mathbf{p}(x_{t+1}) = \bigodot_i^K \hat{\mathbf{g}}_i(t+1).
\end{equation}
Further explanation and detail is provided in Appendix~\ref{sec:appendix_path_integration}.

\subsection{Mapping the model to the HF}
\label{sec:entireHF}

\begin{figure}[t]
   \centering
   \floatbox[{\capbeside\thisfloatsetup{capbesideposition={right,top},capbesidewidth=7cm}}]{figure}[\FBwidth]
   {\caption{\textbf{Schematic of proposed attractor model.} In MEC, the $\mathbf{g}_i$ are residue representations in grid modules, and \textbf{c} encodes a context label. Input of velocity estimate $\mathbf{q}(\mathbf{v})$ can produce path integration in grid modules via binding, denoted by $\odot$. In HC, \textbf{p} represents contextualized  place. Binding serves two roles in the MEC/HC interaction (symbolized by bidirectional arrows): {\it a)} factorizing \textbf{p} into $\mathbf{g}_i$'s, and {\it b)} generating an update of \textbf{p} from the $\mathbf{g}_i$'s, for example, after path integration.  In LEC, \textbf{s} represents sensory input, interacting with \textbf{p} through a learned heteroassociative projection.}   \label{fig:infoflow}}
   {\includegraphics[width=0.95\linewidth]{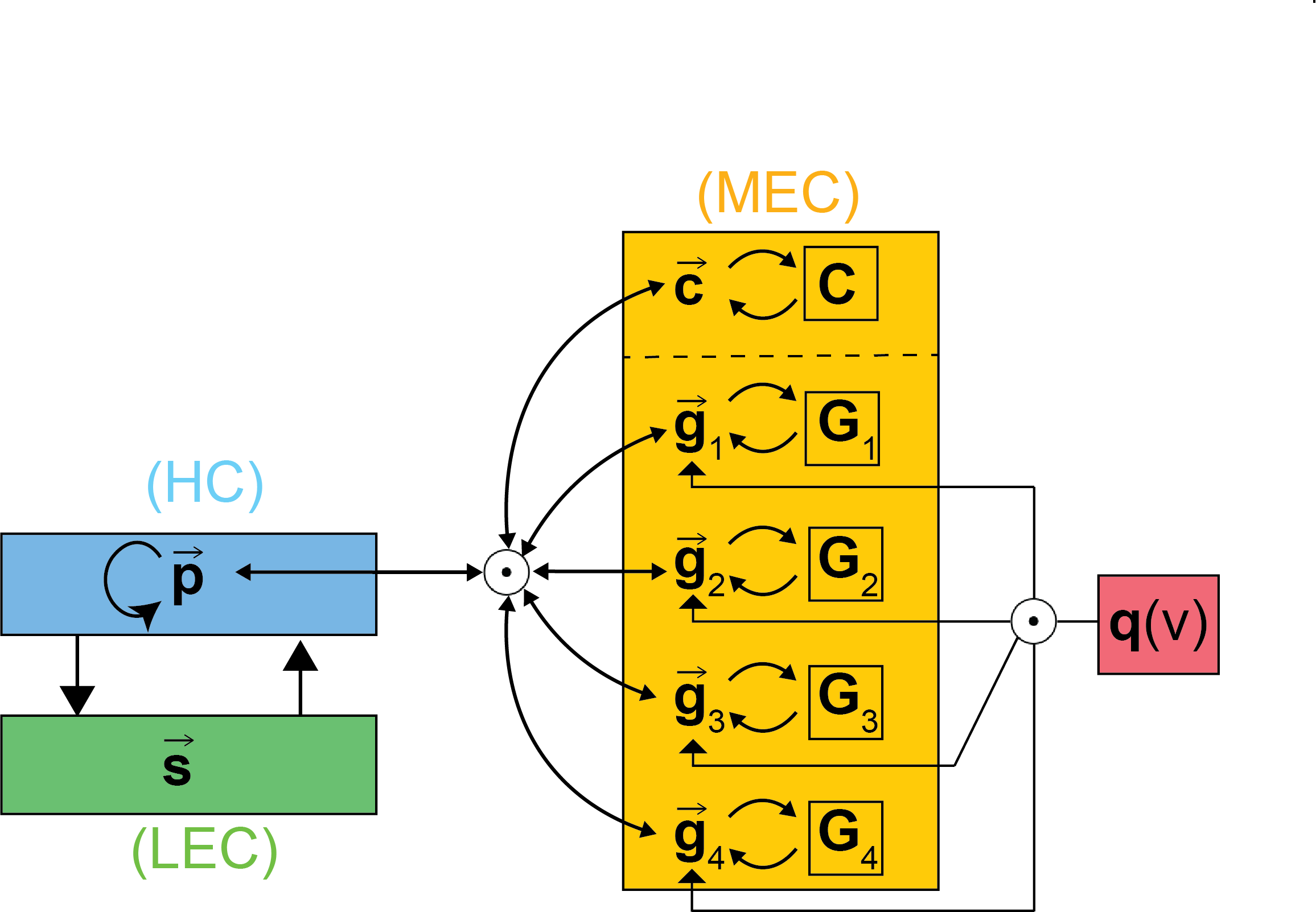}}
\end{figure}

Although it is not obvious how the components of our normative model should map to the anatomical architecture of HF, we make one proposal as shown in Figure~\ref{fig:infoflow}.
The memory networks for residue representations $\mathbf{\hat{g}}_i$ correspond to grid modules in MEC.
Similar to the grid modules, a module for context can be added to the architecture, such as a tag for the identity of a specific environment, with the recurrent synapses $\mathbf{C}$ storing tags of different environments. 

The context neurons could correspond to the non-grid entorhinal cells, which can contain local, non-spatial information about the environment \cite{latuske2018hippocampal}.
The vector $\mathbf{p}(x_t)$ can be linked to place cells in hippocampus. 
Internal HC circuitry can either buffer the input as in Eq.~(\ref{eq:reso}) or allow it to be updated dynamically according to the MEC input (Section \ref{sec:path_integration}). 
The mutual interactions between HC and MEC grid modules require projections between these structures. The binding operations that these interactions involve according to Eq.~(\ref{eq:reso}) are hypothesized to be implemented by nonlinear interactions between dendritic inputs in HC and MEC neurons. 
 
The model also assumes the ability for sensory cues to provide the initialization signal of the cognitive map, represented by $\mathbf{s}$ in Figure~\ref{fig:infoflow}. For completeness, we make the basic assumption that heteroassociative memories are formed by the brain that link sensory cues to the place cell representations $\mathbf{p}$ (Section \ref{sec:heteroassociation_exps}). This process would require the system to generate a new context vector $\mathbf{c}$ and initialize the cognitive map to a default location in order to learn about new environments. We show that through even a simple heteroassociative mechanism, our modular attractor network can robustly retrieve sensory memories and even protect its compositional structure.

\section{Coding properties of the model}

\subsection{RNS representations have exponential coding range}

The compositional RNS vector representation Eq.~(\ref{eq:rnsvecrep}) can encode a coding range of $M$ values using a total of $n$ component patterns for representing the residue of individual modules. The scaling of the coding range is exponential in the number of moduli, $K$, since if each module has $\mathcal{O}(m)$ patterns, and the co-prime condition is satisfied, the scaling of the coding range is $\mathcal{O}(m^K)$. This recovers the expressivity argued by~\cite{fiete2008grid,mathis2012resolution}. 

More generally, it is also exponential in the number of component patterns, $n$. The optimal coding range is given by the best partition of $n$ into a set of positive $\{ m_i \}$. This optimization is identical to that of finding the maximum order of an element in the group of permutations $S_n$, because the maximum order can be found by finding the longest cycle. The scaling of this value in $n$ is characterized by Landau's function $f(n)$, which is known to converge to $\text{exp}(\sqrt{n \ \text{ln} \ n})$ as $n \to \infty$~\cite{landau1903maximalordnung}. Figure~\ref{fig:scaling_noise}A illustrates how Landau's function is the upper bound to what is achievable for any fixed number of moduli ($K$).

Though other kinds of representations can achieve an exponential coding range, the advantage of the compositional encoding of Eq.~(\ref{eq:rnsvecrep}) comes from the fact that the binding operation implements carry-free vector addition (our fourth principle). This enables updates of the encoded value without requiring further transformations such as decoding, facilitating tasks such as path integration (Section \ref{sec:path_integration}, Appendix \ref{sec:appendix_sequences}). Binary representations, by contrast, have exponential coding range but require carry-over operations to implement.

\subsection{The modular attractor network has superlinear coding range}
\label{sec:empirical_scaling}

\begin{figure}[t]

    \centering
    \includegraphics[width=0.9\linewidth]{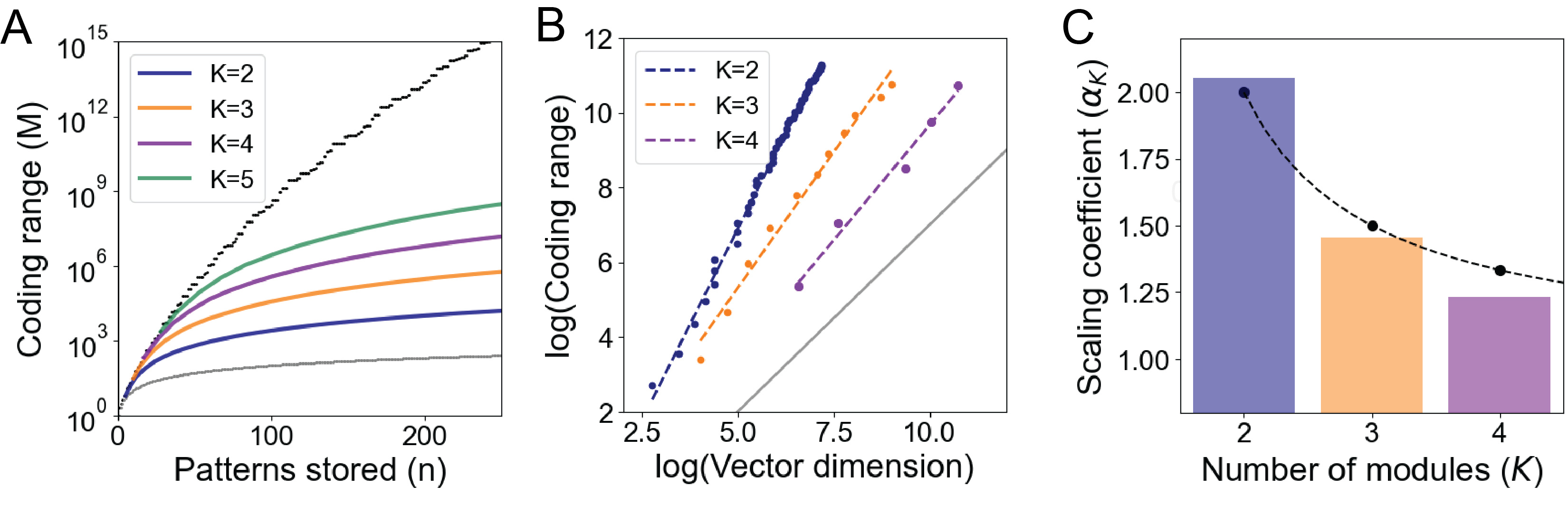}
    \caption{\textbf{Residue number systems, combined with a modular attractor network (resonator network), result in a new kind of attractor neural network with favorable scaling for a large combinatorial range.} \textbf{A)} Number of encoding states, $M$, grows rapidly in the number of modules, up to a maximum established by Landau's function (black dots). \textbf{B)} Coefficient of coding range, M, scales roughly as $\mathcal{O}(D^{\alpha_K})$, depending on the number of moduli, $K$, but with $\alpha_K > 1$.  \textbf{C)} Estimation of scaling from slopes of linear regression (fit to log-log scale). Higher values of $K$ require a higher dimension to achieve a particular coding range; empirical values are close to $\alpha_K = \frac{K}{K-1}$.
    }
    \label{fig:scaling_noise}
\end{figure}
The exponential scaling of the coding range of the RNS representation is a prerequisite to obtain a large coding range with the attractor network that has to perform computations on this representation, such as input denoising, working memory, and path integration.  
To estimate the scaling of the coding range in the proposed attractor network (Eq.~\ref{eq:reso}), we study the critical dimension for which the grid modules converge with high probability. Specifically, we empirically estimate the minimum dimension required to retrieve an arbitrary RNS representation with high probability, given a maximum number of iterations (Figure~\ref{fig:scaling_noise}B). Remarkably, we find that the number of component patterns $n$ that can be stored is superlinear in the pattern dimension $D$; empirically $\mathcal{O}(D^{\alpha})$ for some $\alpha \geq 1$. For 2, 3, and 4 moduli, $\alpha \approx 2.05, 1.45$ and $1.23$, respectively (Figure~\ref{fig:scaling_noise}C).

These empirical scaling laws are consistent with a simple information-theoretic calculation (Appendix~\ref{sec:scaling_infotheory}).
The minimal amount of bits to be stored for the entire RNS vector encoding scheme is of order $\mathcal{O}(M \ \text{log} \ M)$, and the number of synapses in the attractor network is $\mathcal{O}(D\sqrt[K]{M})$. If one makes the cautious assumption of a capacity per synapse of $\mathcal{O}(1)$, the leading order for the coding range $M$ is $\mathcal{O}(D^{\alpha})$, with $\alpha=\frac{K}{K-1}$.  

Note that while the coding range increases with the number of moduli ($K$) for the RNS representation, the superlinear scaling coefficient $\alpha_K$ decreases with $K$ for the modular attractor network, reaching maximum superlinearity at the smallest value $K=2$. This reversal is caused by the fact that increasing $K$ decreases the number of synapses, i.e., the memory resource in the attractor network.  

\subsection{Robust error correction}
\label{sec:noise_robustness}

\begin{figure}[h]

    \centering
    \includegraphics[width=0.9\linewidth]{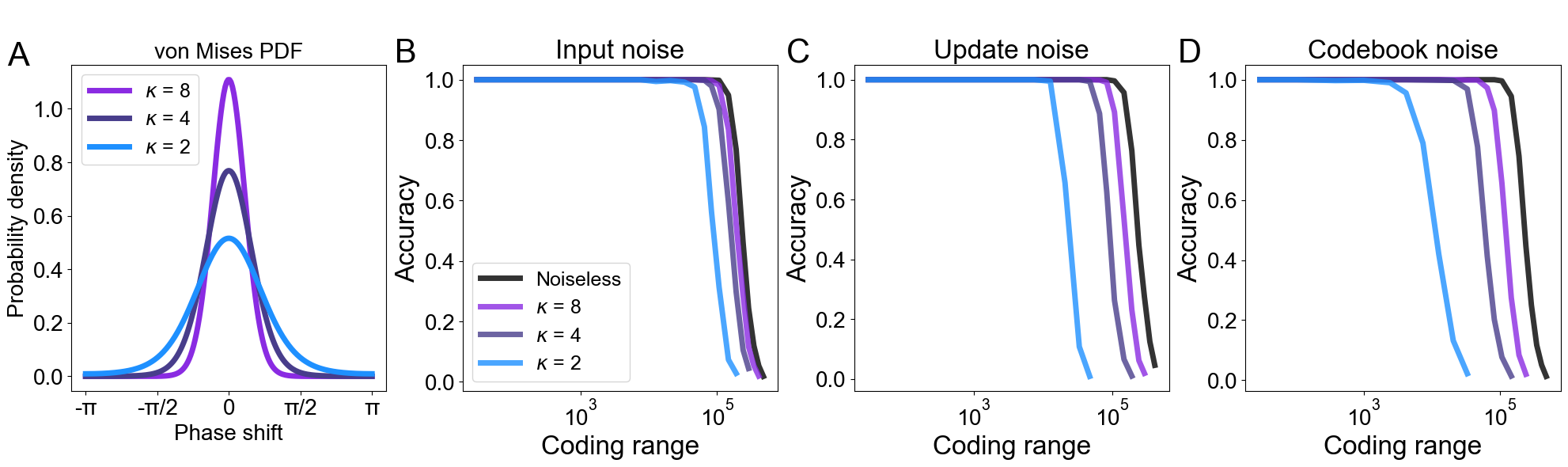}
    \caption{\textbf{Recovery of encoded positions is robust to various sources of noise}. \textbf{A)} Visualization of the von Mises weight distribution. Note that the magnitude of the noise is inversely proportional to $\kappa$, and that the variance of the phase perturbation is much larger than the distance between the discrete states of phasors. \textbf{B-D)} Visualizations of accuracy as a function of coding range and $\kappa$ for three separate cases: input noise (B), update noise (C), and codebook noise (D). Cases are shown in order of increasing difficulty. The resonator network maintains perfect accuracy up to a point, after which accuracy decays at an earlier point than the noiseless dynamics (black curve).}
    \label{fig:error_int}
\end{figure}
In addition, we evaluate the robustness of our attractor model to noise. Because the RNS representations are composed of phasors, which are circular variables, we sample noise from a von Mises distribution with two parameters: mean ($\mu = 0$) and concentration pattern $\kappa$ (Figure~\ref{fig:error_int}A). Higher $\kappa$ values imply less noise; the distribution approximates a Gaussian with variance $1/\kappa$ for large $\kappa$. 

We consider three cases: noisy input patterns, noise added to each time step, and noisy weights corruptions of patterns in $\mathbf{G}_i$ (Appendix~\ref{sec:error_correction_methods}). The empirical accuracy of recall varies depending on the type of corruption applied (Figure~\ref{fig:error_int}A). We find that for a given dimension $D$ (in this case, $1024$), increasing noise decreases the maximum coding range that can be decoded with high accuracy (Figure~\ref{fig:error_int}B-D). For a fixed noise level, the high-accuracy coding range is largest for input noise, followed by update noise and codebook noise. It is perhaps not surprising that codebook noise has the worst coding range, given that noise added to every stored pattern compounds across the dynamics. Fortunately, the demonstrated robustness to input noise enables sensory patterns to be denoised via heteroassociation (Section~\ref{sec:heteroassociation_exps}).

\subsection{Interpolation between patterns enables continuous path integration}

\begin{figure}[h]

    \centering
    \includegraphics[width=0.9\linewidth]{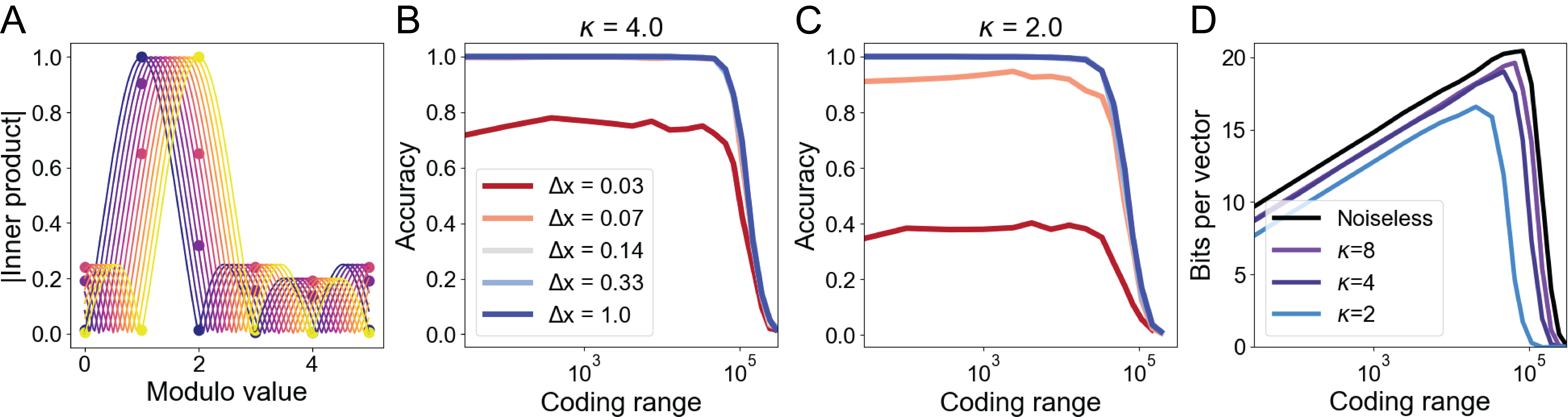}
    \caption{\textbf{Continuity of attractor landscape enables sub-integer decoding and path integration.} \textbf{A)} Visualization of interpolation between two integer states. The position of the fractional value can be estimated by fitting a periodic sinc function (Appendix~\ref{sec:residue_kernel}) based on the inner products with integer codebooks (visualized in dots), then finding the location of the peak. \textbf{B, C)} Sub-integer states can be be decoded, up to a precision set by the noise level. Note that in both cases, sub-integer decoding can be just as accurate as integer decoding for the same range, even though the sub-integer decoding problem is strictly harder. Even $\kappa = 4$ is sufficient to achieve accuracy within a precision of $\Delta x = 0.07$, but for higher noise ($\kappa = 2$), the precision is worse. \textbf{D)} The best spatial precision (in bits) that can be decoded for a fixed noise level. Less noise achieves both a higher coding range and higher information content per vector.}
    \label{fig:continuity_error}
\end{figure}

In general, there is a sharp difference between point and line attractors. In our attractor model, the RNS representations of integer values are stored as discrete fixed points. Nevertheless, the attractor network also converges to states that represent non-integer values that are not explicitly stored. In other words, the network smoothly interpolates to points on a manifold of states that represent integer and non-integer values encoded by (\ref{eq:rnsvecrep}); 
Figure~\ref{fig:continuity_error}A provides a visualization, showing that the kernel induced by inner product operations retains graded similarity for sub-integer shifts. This kernel enables the modular attractor network to settle to fixed points that correspond to interpolations between integers, and for sub-integer positions to be decoded. 

The resolution of decoding is fundamentally limited by the signal to noise ratio. Even so, we find that, up to a fixed noise level, the accuracy regimes of integer decoding and sub-integer decoding coincide. This property enables sub-integer shifts to be encoded within the states of the network, which, as we will show, results in stable, error-correcting path integration (Section \ref{sec:path_integration}). We quantify the gain in precision in terms of the bits of information that can on average be reconstructed from a vector (Figures ~\ref{fig:continuity_error}D, Appendix~\ref{sec:error_correction_methods}). Notably, even a moderate noise level of $\kappa=8$ is sufficient to achieve nearly the same information content as in the noiseless case.

\subsection{Triangular frames in 2D maximize spatial information}
\label{sec:hex_main}
In two-dimensional open field environments, grid cells have firing fields arranged in a hexagonal lattice~\cite{hafting2005microstructure}. Work in theoretical neuroscience shows the optimality of this lattice for 2D environments in terms of spatial information~\cite{wei2015principle,mathis2015probable,anselmi2020computational}. However, the presence of hexagonal firing fields raises a puzzle for residue number systems. Although a crucial property of a RNS is the carry-free property, most implementations of RNS will not perform carry-free updates within a module in non-Cartesian coordinate systems. This generally occurs because the updates of different coordinates must interact due to non-orthogonality. 

We resolve this issue by showing how to implement a version of vector binding of multiple coordinates in a triangular `Mercedes-Benz' frame that enables carry-free hexagonal coding. Furthermore, we provide a combinatoric argument for the optimality of triangular \textit{frames} for $\mathbb{R}^2.$ (A frame is a spanning set for a vector space in which the basis vectors need not be linearly independent.) Our argument relies on the combinatorics of residue numbers, and so for the first time gives an explanation of why the coexistence of RNS and hexagonal codes is optimal.

\begin{wrapfigure}[17]{R}{0.45\textwidth} 
    \includegraphics[width=\linewidth]{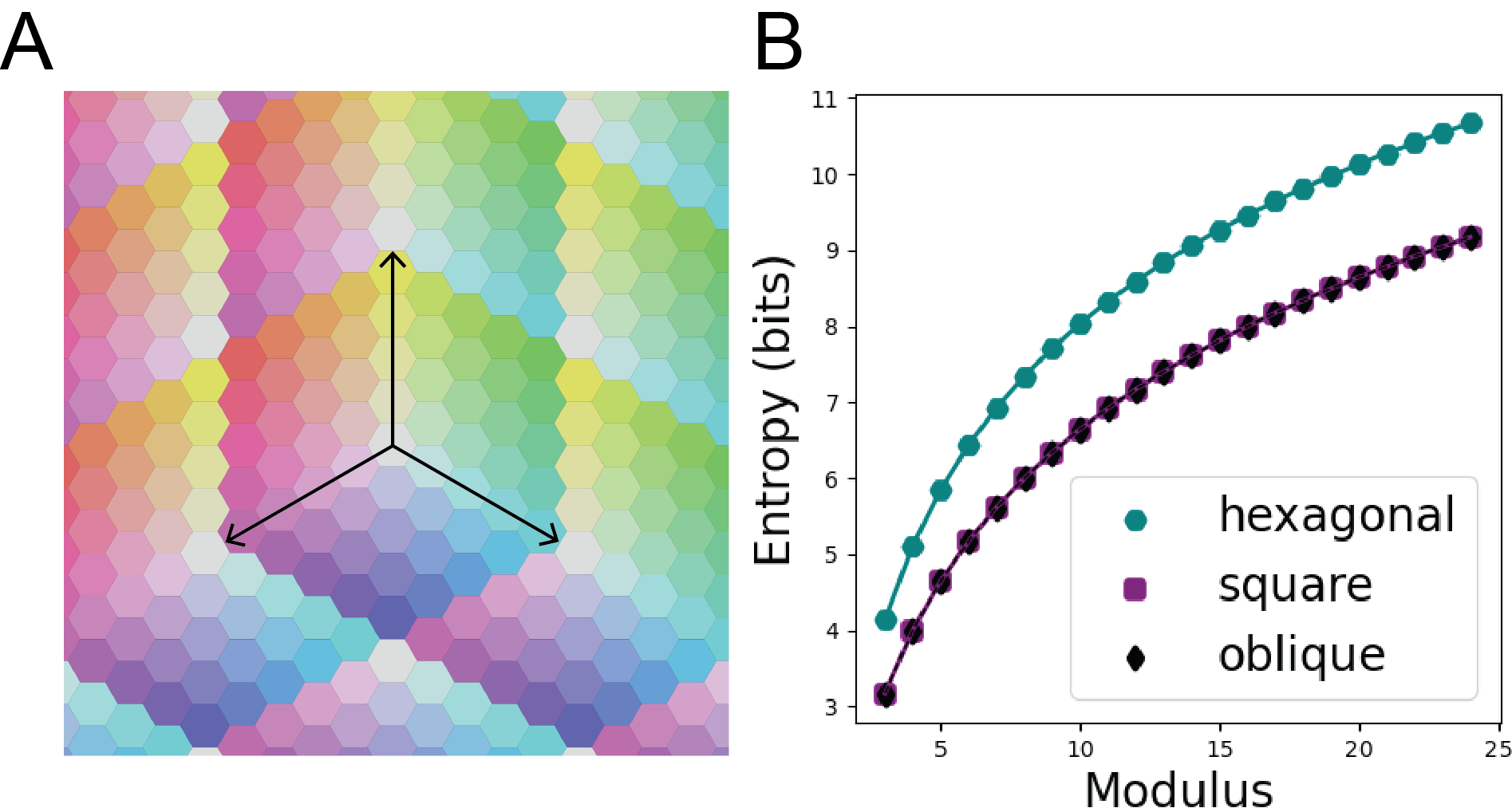}
    \vspace{-0.5cm}
    \caption{\textbf{Hexagonal coding improves spatial resolution.} \textbf{A)} Voronoi tessellation for $m=5$\@. Each distinct color corresponds to a unique codeword in $\mathbb{C}^{D}$\@. Black arrows show the coordinate axes of the triangular `Mercedes-Benz' frame in 2D. \textbf{B)} Hexagonal lattices have higher entropy than square lattices, allowing each state to carry higher resolution in its spatial output.}
    \label{fig:hexagonal}
\end{wrapfigure}

To form a hexagonal tiling of 2D position requires two steps: first, projection into a $3$-coordinate frame, and second, choosing phases such that simultaneous, equal movements along all three frames cancel out (Appendix~\ref{sec:hex_frame}). The resulting Voronoi tessellation for different states is pictured in Figure~\ref{fig:hexagonal}A. This encoding enables higher spatial resolution in terms of the number of discrete states: $3m^2-3m+1$ for triangular frames, versus $m^2$ for Cartesian frames. This increased expressivity results in a higher entropy) code for space (Figure~\ref{fig:hexagonal}B). It also results in both a periodic hexagonal kernel and the individual grid response fields being arranged in a hexagonal lattice (Figure~\ref{fig:path_integration}C).
\leavevmode

Prior models achieved hexagonal lattices either by circularly symmetric receptive fields (e.g.,~\cite{fuhs2006spin,burak2009accurate}) arranged on a periodic rectangular sheet or by distorting a square lattice into an oblique one (e.g.,~\cite{chandra2023high,mosheiff2019velocity}). Importantly, oblique lattices have the combinatorial complexity as the square grid and, unlike the construction described above, they do not achieve the same level of spatial resolution (Figure~\ref{fig:hexagonal}B).

\section{Testing functionalities of the model}

\subsection{Robust path integration}
\label{sec:path_integration}
\begin{figure}[h]
    \centering
    \includegraphics[width=\linewidth]{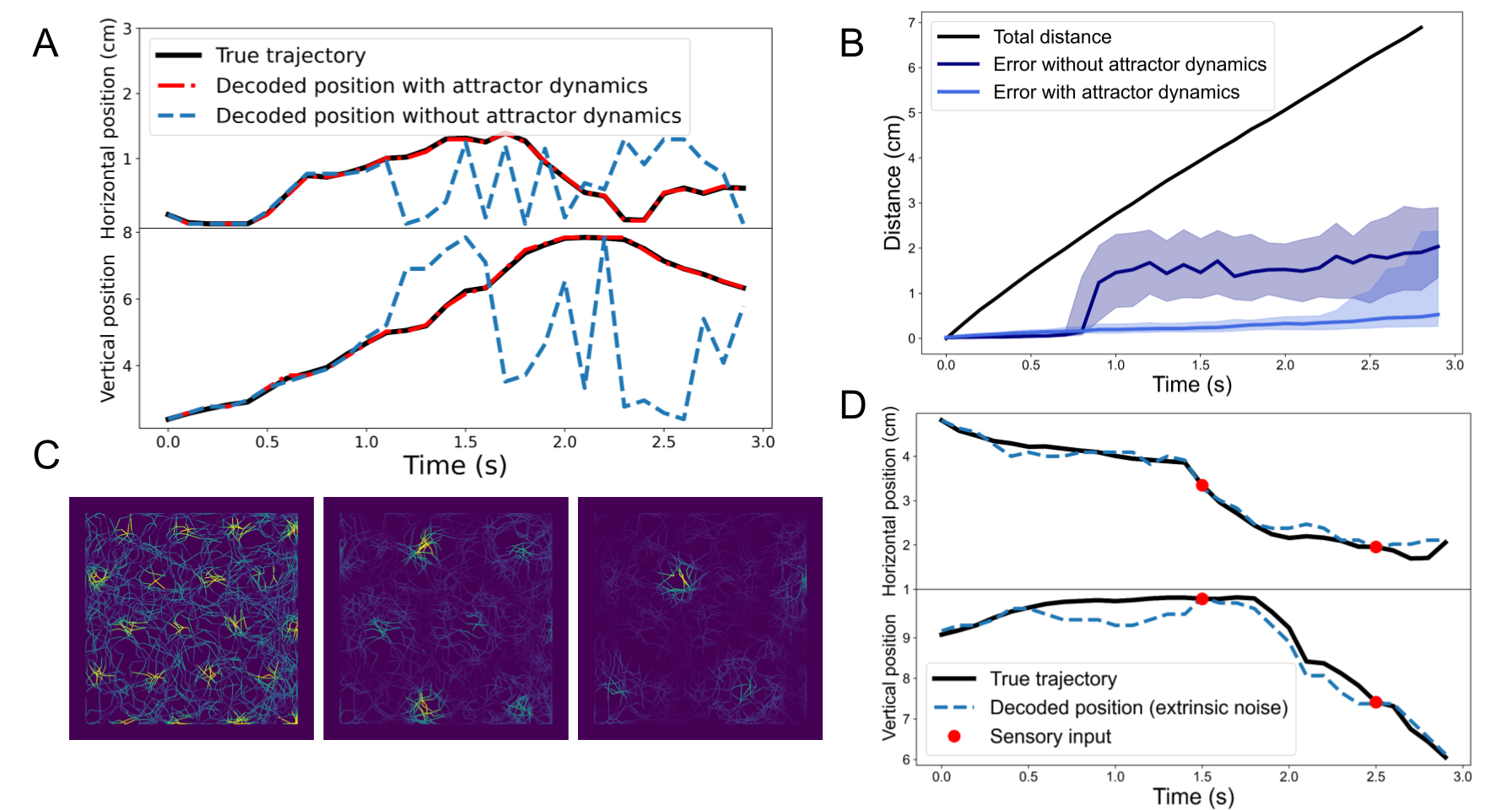}
    \caption{\textbf{Velocity shift mechanism enables robust path integration.} \textbf{A)} Example of path integration of a 2D trajectory in the case of intrinsic input noise on the place cell representation. The grid cell modules correct the noise that would otherwise induce drift after a short period of time. \textbf{B)} Path integration results averaged over multiple trajectories in the case of intrinsic input noise on the place cell representation. Grid cell modules limit noise accumulation along the trajectory.  Solid lines report the median error over $100$ trials, with shaded intervals reporting $25^{\text{th}}$ and $75^{\text{th}}$ percentile. \textbf{C)} Simulated trajectory, along which colors represent the similarity between the $\mathbf{g}_i$ of different modules and vectors representing each position in the environment. 
    We see hexagonal response fields, similar to those obtained from single unit recordings of MEC. \textbf{D)} Sensory patterns (symbolized by red dots), representing visual cues, are associated to positions in the environment. Presentation of visual cues helps correct drifted positions due to extrinsic noise.}
    \label{fig:path_integration}
\end{figure}

Given the ability of the attractor model to update its representation of position from velocity inputs, along with its ability to represent continuous space, we evaluate its ability to perform path integration in the presence of noise. We simulate trajectories based on a statistical model for generating plausible rodent movements in an arena~\cite{raudies2012modeling,banino2018vector}, and we update grid cell and place cell state vectors according to Equations \ref{eqn:update-step} and \ref{eqn:place-update}, respectively. 

To evaluate the robustness of the model to error (Appendix~\ref{sec:appendix_path_integration}), we consider both sources of extrinsic noise (e.g., mis-representations of velocity information), and intrinsic noise (e.g., due to noise in weight updates). The robustness of our model to intrinsic noise is tested by comparing our results to the estimated trajectories obtained without the correction by the MEC modules (Figure~\ref{fig:path_integration} A and B). We find that our model strongly limits noise accumulation along the trajectory and allows highly accurate integration for a longer period of time (Figure~\ref{fig:path_integration}A). Consistent with our previous experiments on noise robustness (Figure~\ref{fig:error_int}), we find strong robustness to intrinsic noise, and that extrinsic noise results in progressive drift of estimated position.

We visualize the response fields in different modules and find hexagonal lattices with a module dependent scaling (Figure~\ref{fig:path_integration}C, Appendix~\ref{sec:path_integration}). In addition, we show that tethering to external cues (e.g., visual inputs), can significantly increase the accuracy of the attractor network. To study this, we associate visual cues to corresponding patches see Section~\ref{sec:heteroassociation_exps}) and observe that integration of information from sensory visual inputs succeeds in correcting drift due to extrinsic noise (Figure~\ref{fig:path_integration}D).

\subsection{Denoising sensory states via a heteroassociative memory}
Finally, we describe a simple extension to our model, in which sensory patterns are fed from the lateral entorhinal cortex (LEC) to update the hippocampal state. This is consistent with theories of memory suggesting that LEC provides the content of experiences to hippocampus~\cite{manns2006evolution}, as well as neuroanatomical evidence~\cite{knierim2014functional}. Although the structure of the representations of those sensory patterns is unknown, it is theorized that HF is critical to sensory pattern completion \cite{teyler1986hippocampal}.

Consistent with this function, recent work~\cite{sharma2022content,chandra2023high} has proposed that a heteroassociative scaffold connects sensory patterns to hippocampal activity, allowing robust denoising of sensory states. Though the main focus of our normative model is not sensory denoising, we show that a simple extension to our model (Appendix~\ref{sec:heteroassociation_methods}) robustly retrieves noisy pattern even under high levels of corruption (Figures ~\ref{fig:heteroassociation}A and B). In Appendix~\ref{sec:appendix_sequences}, we also discuss how this capacity for generalization can serve as a model for sequence retrieval, showing some preliminary experiments.

In addition to robust denoising of single patterns, our model is also well-equipped to deal with compositions of sensory patterns. Two situations are worth emphasizing: first, we can often unmix multiple sensory states corresponding to a sum of patterns, because the compositional structure of binding between grid modules ``protects'' the items in summation (Figure~\ref{fig:heteroassociation}C). This differentiates our model from other heteroassociative memories, in which sums of patterns would have multiple equally valid yet incompatible decodings. Second, the context vector modules allow preservation of different sensory information for different environments (Figure~\ref{fig:heteroassociation_context}).

\label{sec:heteroassociation_exps}
\begin{figure}[h]
    \centering
    \includegraphics[width=0.9\linewidth]{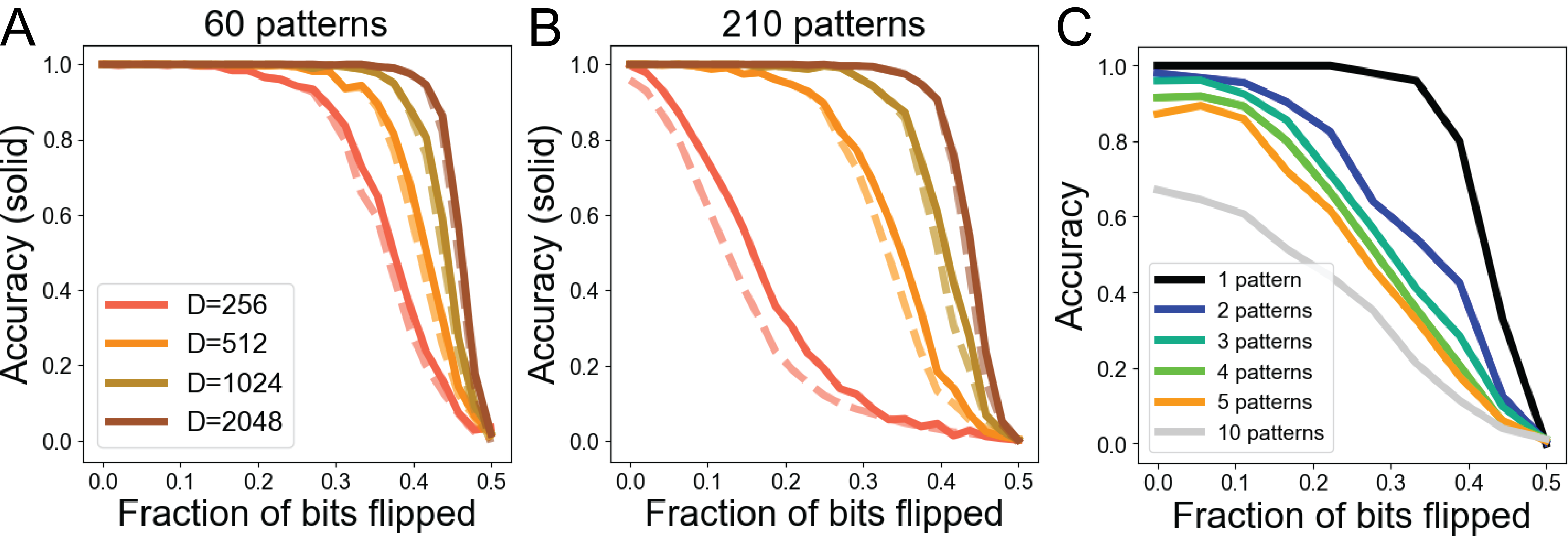}
    \caption{\textbf{Heteroassociation enables recovery of sensory patterns under noise and superposition.} \textbf{A)} Accuracy for denoising $60$ different random binary patterns for different vector dimension $D$. The dotted line is the average similarity between the decoded and ground truth patterns. \textbf{B)} Same experiment as in panel A, but with $210$ different possible binary patterns. The accuracy is lower on average. \textbf{C)} Accuracy for denoising multiple patterns from a single input. This task is especially challenging, because sums of patterns combined in this way interfere with each other in retrieval (a phenomenon known as cross-talk noise). However, the compositional structure of our modular attractor network enables multiple patterns to be decoded with high probability.}
    \label{fig:heteroassociation}
\end{figure}

\section{Discussion}
\label{sec:discussion}
We propose a normative model of a cognitive map for the hippocampal formation in the mammalian brain. The core principle of the model is a compositional representation of space that achieves a superlinear coding range, which is expressed by a compact, multi-module attractor network. The compositional mechanism of vector binding provides generalization to multiple spatial dimensions, contextualization, and path integration. This binding mechanism builds on prior work proposed in the field of hyperdimensional computing and vector symbolic architectures \cite{kanerva2009hyperdimensional,gayler2004vector,plate1992holographic,dumont2022model,kymn2023computing} --- and goes beyond it to develop a specific algorithmic hypothesis about structured operations in HF. Our analyses and experiments confirm that the model can achieve important functions of the hippocampal formation and explains experimental observations, such as hexagonal grid cells, place cells, and remapping phenomena.

The proposed model contributes to, and greatly benefits from, existing work in theoretical neuroscience on residue number systems~\cite{fiete2008grid,sreenivasan2011grid}, continuous attractor network models of grid cells~\cite{fuhs2006spin,zhang1996representation,fiete2008grid}, and the optimality of hexagonal representations in 2D~\cite{mathis2012resolution,mathis2015probable}. It remains intriguing that biology organized grid cells into multiple discrete modules, rather than pooling all resources into a single module attractor network. This puzzle raises an opportunity for normative models to explain the organization of grid cells into multiple modules. More recent work has focused on the problem of coordinating representations across multiple modules~\cite{mosheiff2017efficient,mosheiff2019velocity,kang2019geometric,agmon2020theory,chandra2023high}, and large scale recordings of HF \cite{waaga2022grid} may provide new opportunities to evaluate predictions of these different ideas.

Our approach starts from principles of space encoding, in particular, the requirement of compositionality. This strategy is complimentary to, but different from, investigations of the emergence of place and grid cells in artificial neural networks (e.g.,~\cite{banino2018vector,cueva2018emergence,whittington2022disentanglement,dorrell2022actionable,sorscher2023unified,schaeffer2024self, stachenfeld2017hippocampus,whittington2020tolman,chen2022predictive}). These approaches show optimality of biological response features under the model assumptions, such as ANN properties, network architecture, training objective and protocol.
Here, we emphasize the role of multiplicative binding, a primitive that is typically difficult to have emerge in an ANN setting. Early suggestions for realizing conjunctive binding already ventured outside the framework of ANNs~\cite{SmolenskyTensor1990, plate1992holographic}. A simple extension of ANNs are sigma-pi neurons~\cite{FeldmanConnectionist1982, mel1989sigma} that can implement vector binding~\cite{plate2000randomly}. Recent work amplifies the view that full conjunctive binding would be a useful inductive bias to augment deep learning architectures~\cite{goyal2022inductive}, and various augmentations of ANNs with dedicated binding mechanisms have been proposed~\cite{danihelka2016associative,greff2020binding,ganesan2021learning,smolensky2022neurocompositional}.  

Our model has obvious limitations. 
Our attractor model for the cognitive map is still a high-level abstraction of spiking neural circuits in the hippocampal formation.  In particular, the phasor states in the model are one linear transform removed from vectors that describe neural population activity. Thus, the mapping between model and neurobiological mechanisms is not straight-forward, a disadvantage that can be addressed by switching to other encoding schemes, such as sparse real or complex vectors, e.g.,~\cite{laiho2015high}, for which conjunctive binding operations have been proposed~\cite{frady2021variable}. 
Although the model is more comprehensive than typical normative models, which usually focus on a single computation, it is far from covering the many other functional cell types observed in the hippocampal formation or contextual modulations observed during remapping.
In addition, the current model includes learning only in the heteroassociative projection to LEC. Most observations regarding plasticity in HF are not captured, i.e., signals from reward, or eligibility traces. Finally, our assumptions about inputs to HF from the sensory pathway are rather simplifying and primarily intended as a proof of concept.

The purpose of the model express the fundamental principles of a compositional cognitive map, permitting testable predictions: First, at the biophysical level, the model predicts multiplicative interactions between dendritic inputs providing the conjunctive binding operation. Though some evidence of MEC-LEC binding exists~\cite{latuske2018hippocampal}, our attractor model also predicts binding between MEC modules. Second, the model predicts relatively fixed attractor weights between place and grid cells, and more plasticity from the hippocampus to sensory observations. 
Third, we predict that causal perturbations of one grid module can affect the states of other grid modules without involvement of the hippocampus, in a direction that is self-consistent with the update of the attractor state.

We believe that the proposed modeling approach and the specific attractor model have broader applications in neuroscience. The proposed attractor network can also model generative models in sensory systems to implement analysis by synthesis postulated in perception. Further, there is a intriguing connection between the proposed phasor models and spiking neural networks~\cite{frady2019robust}, which could yield normative models with spiking neurons, potentially implementable on neuromorphic hardware at large scale that can lead to further quantitative predictions.

\newpage
\section*{Acknowledgments}
The work of CJK was supported by the Department of Defense (DoD) through the National Defense Science \& Engineering Graduate (NDSEG) Fellowship Program. The
work of SM was carried out as part of the ARPE program of ENS Paris-Saclay. The work of DK and BAO was supported in part by Intel’s THWAI program. The work of CJK and BAO was supported by the Center for the Co-Design of Cognitive Systems (CoCoSys), one of seven centers in JUMP 2.0, a Semiconductor Research Corporation (SRC) program sponsored by DARPA. DK has received funding from theEuropean Union’s Horizon 2020 research and innovation programme under the Marie Sklodowska-Curie grant agreement No 839179. FTS discloses support for the research of this work from NIH grant 1R01EB026955-0.

{\small \printbibliography}

@article{frady2021variable,
  title={Variable binding for sparse distributed representations: Theory and applications},
  author={Frady, Edward Paxon and Kleyko, Denis and Sommer, Friedrich T},
  journal={IEEE Transactions on Neural Networks and Learning Systems},
  volume={34},
  number={5},
  pages={2191--2204},
  year={2023},
}

@inproceedings{frady2022computing,
  title={Computing on functions using randomized vector representations (in brief)},
  author={Frady, E Paxon and Kleyko, Denis and Kymn, Christopher J and Olshausen, Bruno A and Sommer, Friedrich T},
  booktitle={Annual Neuro-Inspired Computational Elements Conference (NICE)},
  pages={115--122},
  year={2022}
}

@InProceedings{gayler2004vector,
	author = {R. W. Gayler},
	title = {Vector Symbolic Architectures Answer {J}ackendoff's Challenges for Cognitive Neuroscience},
	booktitle = "{Joint International Conference on Cognitive Science (ICCS/ASCS)}",
	year = "2003",
	pages = "133--138",
}

@inproceedings{plate1991holographic,
  title={Holographic Reduced Representations: Convolution Algebra for Compositional Distributed Representations.},
  author={Plate, Tony and others},
  booktitle = {International Joint Conference on Artificial Intelligence (IJCAI)},
  pages={30--35},
  year={1991}
}

@BOOK{smith2011spectral,
	AUTHOR = "Julius O. Smith",
	TITLE = {Spectral Audio Signal Processing},
	YEAR = {2011},
}

@article{wills2005attractor,
  title={Attractor dynamics in the hippocampal representation of the local environment},
  author={Wills, Tom J and Lever, Colin and Cacucci, Francesca and Burgess, Neil and O'Keefe, John},
  journal={Science},
  volume={308},
  number={5723},
  pages={873--876},
  year={2005},
  publisher={American Association for the Advancement of Science}
}

@article{fuhs2006spin,
  title={A spin glass model of path integration in rat medial entorhinal cortex},
  author={Fuhs, Mark C and Touretzky, David S},
  journal={Journal of Neuroscience},
  volume={26},
  number={16},
  pages={4266--4276},
  year={2006},
  publisher={Soc Neuroscience}
}

@article{burak2009accurate,
  title={Accurate path integration in continuous attractor network models of grid cells},
  author={Burak, Yoram and Fiete, Ila R},
  journal={PLoS Computational Biology},
  volume={5},
  number={2},
  pages={e1000291},
  year={2009},
  publisher={Public Library of Science San Francisco, USA}
}

@article{behrens2018cognitive,
  title={What is a cognitive map? {O}rganizing knowledge for flexible behavior},
  author={Behrens, Timothy EJ and Muller, Timothy H and Whittington, James CR and Mark, Shirley and Baram, Alon B and Stachenfeld, Kimberly L and Kurth-Nelson, Zeb},
  journal={Neuron},
  volume={100},
  number={2},
  pages={490--509},
  year={2018},
  
}

@article{agmon2020theory,
  title={A theory of joint attractor dynamics in the hippocampus and the entorhinal cortex accounts for artificial remapping and grid cell field-to-field variability},
  author={Agmon, Haggai and Burak, Yoram},
  journal={Elife},
  volume={9},
  pages={e56894},
  year={2020},
  publisher={eLife Sciences Publications, Ltd}
}

@article{kang2019geometric,
  title={A geometric attractor mechanism for self-organization of entorhinal grid modules},
  author={Kang, Louis and Balasubramanian, Vijay},
  journal={Elife},
  volume={8},
  pages={e46687},
  year={2019},
  publisher={eLife Sciences Publications, Ltd}
}

@article{fiete2008grid,
  title={What grid cells convey about rat location},
  author={Fiete, Ila R and Burak, Yoram and Brookings, Ted},
  journal={Journal of Neuroscience},
  volume={28},
  number={27},
  pages={6858--6871},
  year={2008},
  publisher={Soc Neuroscience}
}

@article{sreenivasan2011grid,
  title={Grid cells generate an analog error-correcting code for singularly precise Neural Computation},
  author={Sreenivasan, Sameet and Fiete, Ila},
  journal={Nature Neuroscience},
  volume={14},
  number={10},
  pages={1330--1337},
  year={2011},
  publisher={Nature Publishing Group US New York}
}

@article{anselmi2020computational,
  title={A computational model for grid maps in neural populations},
  author={Anselmi, Fabio and Murray, Micah M and Franceschiello, Benedetta},
  journal={Journal of Computational Neuroscience},
  volume={48},
  pages={149--159},
  year={2020},
}

@inproceedings{garner1959residue,
  title={The residue number system},
  author={Garner, Harvey L},
  booktitle={Western Joint Computer Conference (WJCC)},
  pages={146--153},
  year={1959}
}

@phdthesis{komer2020biologically,
  title={Biologically Inspired Spatial Representation},
  author={Komer, Brent},
  year={2020},
  school={University of Waterloo},
}

@InProceedings{KomerNavigation2020,
	author = {B. Komer and C. Eliasmith},
	title = {Efficient Navigation using a Scalable, Biologically Inspired Spatial Representation},
	booktitle = {Annual Meeting of the Cognitive Science Society (CogSci)},
	year = "2020",
	pages = "1532--1538"
}

@article{waaga2022grid,
  title={Grid-cell modules remain coordinated when neural activity is dissociated from external sensory cues},
  author={Waaga, Torgeir and Agmon, Haggai and Normand, Valentin A and Nagelhus, Anne and Gardner, Richard J and Moser, May-Britt and Moser, Edvard I and Burak, Yoram},
  journal={Neuron},
  volume={110},
  number={11},
  pages={1843--1856},
  year={2022},
  
}

@article{landau1903maximalordnung,
  title={{\"U}ber die maximalordnung der permutationen gegebenen grades},
  author={Landau, Edmund},
  journal={Archiv der Mathematik und Physik},
  volume={3},
  pages={92--103},
  year={1903}
}

@article{mathis2012resolution,
  title={Resolution of nested neuronal representations can be exponential in the number of neurons},
  author={Mathis, Alexander and Herz, Andreas VM and Stemmler, Martin B},
  journal={Physical Review Letters},
  volume={109},
  number={1},
  pages={018103},
  year={2012},
  
}

@article{noest1988discrete,
  title={Discrete-state phasor neural networks},
  author={Noest, Andr{\'e} J},
  journal={Physical Review A},
  volume={38},
  number={4},
  pages={2196},
  year={1988},
  
}

@article{frady2020resonator,
  title={Resonator networks, 1: An efficient solution for factoring high-dimensional, distributed representations of data structures},
  author={Frady, E Paxon and Kent, Spencer J and Olshausen, Bruno A and Sommer, Friedrich T},
  journal={Neural Computation},
  volume={32},
  number={12},
  pages={2311--2331},
  year={2020},
}

@article{mathis2015probable,
  title={Probable nature of higher-dimensional symmetries underlying mammalian grid-cell activity patterns},
  author={Mathis, Alexander and Stemmler, Martin B and Herz, Andreas VM},
  journal={Elife},
  volume={4},
  pages={e05979},
  year={2015},
}

@article{raudies2012modeling,
  title={Modeling boundary vector cell firing given optic flow as a cue},
  author={Raudies, Florian and Hasselmo, Michael E},
  journal={PLoS Computational Biology},
  volume={8},
  number={6},
  pages={e1002553},
  year={2012},
}

@article{stensola2012entorhinal,
  title={The entorhinal grid map is discretized},
  author={Stensola, Hanne and Stensola, Tor and Solstad, Trygve and Fr{\o}land, Kristian and Moser, May-Britt and Moser, Edvard I},
  journal={Nature},
  volume={492},
  number={7427},
  pages={72--78},
  year={2012},
  
}

@article{hafting2005microstructure,
  title={Microstructure of a spatial map in the entorhinal cortex},
  author={Hafting, Torkel and Fyhn, Marianne and Molden, Sturla and Moser, May-Britt and Moser, Edvard I},
  journal={Nature},
  volume={436},
  number={7052},
  pages={801--806},
  year={2005},
  
}

@InProceedings{rahimi2007random,
  title={Random features for large-scale kernel machines},
  author={Rahimi, Ali and Recht, Benjamin},
  booktitle = "{Advances in Neural Information Processing Systems (NeurIPS)}",
  volume={20},
  year={2007}
}

@article{hopfield1982neural,
  title={Neural networks and physical systems with emergent collective computational abilities.},
  author={Hopfield, John J},
  journal={Proceedings of the National Academy of Sciences},
  volume={79},
  number={8},
  pages={2554--2558},
  year={1982},
}

@article{manns2006evolution,
  title={Evolution of declarative memory},
  author={Manns, Joseph R and Eichenbaum, Howard},
  journal={Hippocampus},
  volume={16},
  number={9},
  pages={795--808},
  year={2006},
}

@article{kent2020resonator,
  title={Resonator networks, 2: Factorization performance and capacity compared to optimization-based methods},
  author={Kent, Spencer J and Frady, E Paxon and Sommer, Friedrich T and Olshausen, Bruno A},
  journal={Neural Computation},
  volume={32},
  number={12},
  pages={2332--2388},
  year={2020},
}

@inproceedings{noest1987phasor,
  title={Phasor neural networks},
  author={Noest, Andre},
  booktitle = "{Advances in Neural Information Processing Systems (NeurIPS)}",
  pages={584--591},
  year={1987}
}

@article{kymn2023computing,
  title={Computing with Residue Numbers in High-Dimensional Representation},
  author={Kymn, Christopher J and Kleyko, Denis and Frady, E Paxon and Bybee, Connor and Kanerva, Pentti and Sommer, Friedrich T and Olshausen, Bruno A},
  journal={ArXiv},
  year={2023},
}

@article{zhang1996representation,
  title={Representation of spatial orientation by the intrinsic dynamics of the head-direction cell ensemble: a theory},
  author={Zhang, Kechen},
  journal={Journal of Neuroscience},
  volume={16},
  number={6},
  pages={2112--2126},
  year={1996},
}

@inproceedings{sharma2022content,
  title={Content addressable memory without catastrophic forgetting by heteroassociation with a fixed scaffold},
  author={Sharma, Sugandha and Chandra, Sarthak and Fiete, Ila},
  booktitle={International Conference on Machine Learning (ICML)},
  pages={19658--19682},
  year={2022},
}

@article{chandra2023high,
  title={High-capacity flexible hippocampal associative and episodic memory enabled by prestructured ``spatial'' representations},
  author={Chandra, Sarthak and Sharma, Sugandha and Chaudhuri, Rishidev and Fiete, Ila},
  journal={bioRxiv},
  year={2023},
}

@article{frady2018theory,
  title={A theory of sequence indexing and working memory in recurrent neural networks},
  author={Frady, E Paxon and Kleyko, Denis and Sommer, Friedrich T},
  journal={Neural Computation},
  volume={30},
  number={6},
  pages={1449--1513},
  year={2018},
  publisher={MIT Press One Rogers Street, Cambridge, MA 02142-1209, USA journals-info~…}
}

@article{kleyko2023efficient,
  title={Efficient decoding of compositional structure in holistic representations},
  author={Kleyko, Denis and Bybee, Connor and Huang, Ping-Chen and Kymn, Christopher J and Olshausen, Bruno A and Frady, E Paxon and Sommer, Friedrich T},
  journal={Neural Computation},
  volume={35},
  number={7},
  pages={1159--1186},
  year={2023},
  publisher={MIT Press One Rogers Street, Cambridge, MA 02142-1209, USA journals-info~…}
}

@article{wei2015principle,
  title={A principle of economy predicts the functional architecture of grid cells},
  author={Wei, Xue-Xin and Prentice, Jason and Balasubramanian, Vijay},
  journal={Elife},
  volume={4},
  pages={e08362},
  year={2015},
  publisher={eLife Sciences Publications, Ltd}
}

@article{banino2018vector,
  title={Vector-based navigation using grid-like representations in artificial agents},
  author={Banino, Andrea and Barry, Caswell and Uria, Benigno and Blundell, Charles and Lillicrap, Timothy and Mirowski, Piotr and Pritzel, Alexander and Chadwick, Martin J and Degris, Thomas and Modayil, Joseph and others},
  journal={Nature},
  volume={557},
  number={7705},
  pages={429--433},
  year={2018},
  
}

@article{sorscher2023unified,
  title={A unified theory for the computational and mechanistic origins of grid cells},
  author={Sorscher, Ben and Mel, Gabriel C and Ocko, Samuel A and Giocomo, Lisa M and Ganguli, Surya},
  journal={Neuron},
  volume={111},
  number={1},
  pages={121--137},
  year={2023},
  
}

@article{thomas2021theoretical,
  title={A theoretical perspective on hyperdimensional computing},
  author={Thomas, Anthony and Dasgupta, Sanjoy and Rosing, Tajana},
  journal={Journal of Artificial Intelligence Research},
  volume={72},
  pages={215--249},
  year={2021}
}

@InProceedings{cueva2018emergence,
  title={Emergence of grid-like representations by training recurrent neural networks to perform spatial localization},
  author={Cueva, Christopher J and Wei, Xue-Xin},
  booktitle={International Conference on Learning Representations (ICLR)},
  pages={1--19},
  year={2018}
}

@InProceedings{schaeffer2024self,
  title={Self-supervised learning of representations for space generates multi-modular grid cells},
  author={Schaeffer, Rylan and Khona, Mikail and Ma, Tzuhsuan and Eyzaguirre, Cristobal and Koyejo, Sanmi and Fiete, Ila},
  booktitle = "{Advances in Neural Information Processing Systems (NeurIPS)}",
  volume={36},
  year={2023}
}

@article{whittington2020tolman,
  title={The {Tolman-Eichenbaum} machine: unifying space and relational memory through generalization in the hippocampal formation},
  author={Whittington, James CR and Muller, Timothy H and Mark, Shirley and Chen, Guifen and Barry, Caswell and Burgess, Neil and Behrens, Timothy EJ},
  journal={Cell},
  volume={183},
  number={5},
  pages={1249--1263},
  year={2020},
  
}

@article{stachenfeld2017hippocampus,
  title={The hippocampus as a predictive map},
  author={Stachenfeld, Kimberly L and Botvinick, Matthew M and Gershman, Samuel J},
  journal={Nature Neuroscience},
  volume={20},
  number={11},
  pages={1643--1653},
  year={2017},
  publisher={Nature Publishing Group US New York}
}

@article{chen2022predictive,
  title={Predictive Sequence Learning in the Hippocampal Formation},
  author={Chen, Yusi and Zhang, Huanqiu and Cameron, Mia and Sejnowski, Terrrence},
  journal={bioRxiv},
  year={2022},
  publisher={Cold Spring Harbor Laboratory}
}

@article{mosheiff2019velocity,
  title={Velocity coupling of grid cell modules enables stable embedding of a low dimensional variable in a high dimensional neural attractor},
  author={Mosheiff, Noga and Burak, Yoram},
  journal={Elife},
  volume={8},
  pages={e48494},
  year={2019},
  publisher={eLife Sciences Publications, Ltd}
}

@article{mosheiff2017efficient,
  title={An efficient coding theory for a dynamic trajectory predicts non-uniform allocation of entorhinal grid cells to modules},
  author={Mosheiff, Noga and Agmon, Haggai and Moriel, Avraham and Burak, Yoram},
  journal={PLoS Computational Biology},
  volume={13},
  number={6},
  pages={e1005597},
  year={2017},
  
}

@article{knierim2014functional,
  title={Functional correlates of the lateral and medial entorhinal cortex: objects, path integration and local--global reference frames},
  author={Knierim, James J and Neunuebel, Joshua P and Deshmukh, Sachin S},
  journal={Philosophical Transactions of the Royal Society B: Biological Sciences},
  volume={369},
  number={1635},
  pages={20130369},
  year={2014},
  publisher={The Royal Society}
}

@article{thompson1989place,
  title={Place cells and silent cells in the hippocampus of freely-behaving rats},
  author={Thompson, LT and Best, PJ},
  journal={Journal of Neuroscience},
  volume={9},
  number={7},
  pages={2382--2390},
  year={1989},
  publisher={Soc Neuroscience}
}

@inproceedings{dumont2022model,
  title={A model of path integration that connects neural and symbolic representation},
  author={Dumont, Nicole Sandra-Yaffa and Orchard, Jeff and Eliasmith, Chris},
  booktitle={Proceedings of the Annual Meeting of the Cognitive Science Society},
  volume={44},
  number={44},
  year={2022}
}

@article{latuske2018hippocampal,
  title={Hippocampal remapping and its entorhinal origin},
  author={Latuske, Patrick and Kornienko, Olga and Kohler, Laura and Allen, Kevin},
  journal={Frontiers in Behavioral Neuroscience},
  volume={11},
  pages={253},
  year={2018},
  
}

@article{kurth2023replay,
  title={Replay and compositional computation},
  author={Kurth-Nelson, Zeb and Behrens, Timothy and Wayne, Greg and Miller, Kevin and Luettgau, Lennart and Dolan, Ray and Liu, Yunzhe and Schwartenbeck, Philipp},
  journal={Neuron},
  volume={111},
  number={4},
  pages={454--469},
  year={2023},
  
}

@article{moser2017spatial,
  title={Spatial representation in the hippocampal formation: a history},
  author={Moser, Edvard I and Moser, May-Britt and McNaughton, Bruce L},
  journal={Nature Neuroscience},
  volume={20},
  number={11},
  pages={1448--1464},
  year={2017},
}

@article{eichenbaum2017integration,
  title={On the integration of space, time, and memory},
  author={Eichenbaum, Howard},
  journal={Neuron},
  volume={95},
  number={5},
  pages={1007--1018},
  year={2017},
  
}

@article{schlesiger2015medial,
  title={The medial entorhinal cortex is necessary for temporal organization of hippocampal neuronal activity},
  author={Schlesiger, Magdalene I and Cannova, Christopher C and Boublil, Brittney L and Hales, Jena B and Mankin, Emily A and Brandon, Mark P and Leutgeb, Jill K and Leibold, Christian and Leutgeb, Stefan},
  journal={Nature Neuroscience},
  volume={18},
  number={8},
  pages={1123--1132},
  year={2015},
}

@article{yamamoto2017direct,
  title={Direct medial entorhinal cortex input to hippocampal {CA1} is crucial for extended quiet awake replay},
  author={Yamamoto, Jun and Tonegawa, Susumu},
  journal={Neuron},
  volume={96},
  number={1},
  pages={217--227},
  year={2017},
  
}

@article{greff2020binding,
  title={On the binding problem in artificial neural networks},
  author={Greff, Klaus and Van Steenkiste, Sjoerd and Schmidhuber, J{\"u}rgen},
  journal={arXiv:2012.05208},
  year={2020}
}

@article{constantinescu2016organizing,
  title={Organizing conceptual knowledge in humans with a gridlike code},
  author={Constantinescu, Alexandra O and O’Reilly, Jill X and Behrens, Timothy EJ},
  journal={Science},
  volume={352},
  number={6292},
  pages={1464--1468},
  year={2016},
  publisher={American Association for the Advancement of Science}
}

@article{bellmund2018navigating,
  title={Navigating cognition: Spatial codes for human thinking},
  author={Bellmund, Jacob LS and G{\"a}rdenfors, Peter and Moser, Edvard I and Doeller, Christian F},
  journal={Science},
  volume={362},
  number={6415},
  year={2018},
  publisher={American Association for the Advancement of Science}
}

@inproceedings{ganesan2021learning,
  title={Learning with holographic reduced representations},
  author={Ganesan, Ashwinkumar and Gao, Hang and Gandhi, Sunil and Raff, Edward and Oates, Tim and Holt, James and McLean, Mark},
  booktitle = "{Advances in Neural Information Processing Systems (NeurIPS)}",
  volume={34},
  pages={25606--25620},
  year={2021}
}

@inproceedings{danihelka2016associative,
  title={Associative long short-term memory},
  author={Danihelka, Ivo and Wayne, Greg and Uria, Benigno and Kalchbrenner, Nal and Graves, Alex},
  booktitle={International Conference on Machine Learning (ICML)},
  pages={1986--1994},
  year={2016},
}

@inproceedings{whittington2022disentanglement,
  title={Disentanglement with biological constraints: A theory of functional cell types},
  author={Whittington, James CR and Dorrell, Will and Ganguli, Surya and Behrens, Timothy},
  booktitle={International Conference on Learning Representations (ICLR)},
  year={2022}
}

@inproceedings{dorrell2022actionable,
  title={Actionable neural representations: Grid cells from minimal constraints},
  author={Dorrell, William and Latham, Peter E and Behrens, Timothy EJ and Whittington, James CR},
  booktitle={International Conference on Learning Representations (ICLR)},
  year={2023}
}

@inproceedings{kymn2024compositional,
  title={Compositional Factorization of Visual Scenes with Convolutional Sparse Coding and Resonator Networks},
  author={Kymn, Christopher J and Mazelet, Sonia and Ng, Annabel and Kleyko, Denis and Olshausen, Bruno A},
  booktitle={2024 Neuro Inspired Computational Elements Conference (NICE)},
  pages={1--9},
  year={2024},
  organization={IEEE}
}

@book{plate2003holographic,
  title={Holographic Reduced Representation: Distributed representation for cognitive structures},
  author={Plate, Tony A},
  volume={150},
  year={2003},
  publisher={CSLI Publications Stanford}
}

@article{clarkson2023capacity,
  title={Capacity analysis of vector symbolic architectures},
  author={Clarkson, Kenneth L and Ubaru, Shashanka and Yang, Elizabeth},
  journal={arXiv preprint arXiv:2301.10352},
  year={2023}
}

@article{plate2000randomly,
  title={Randomly connected {Sigma-Pi} neurons can form associator networks},
  author={Plate, Tony A},
  journal={Network: Computation in Neural Systems},
  volume={11},
  number={4},
  pages={321},
  year={2000},
}

@article{kanerva2009hyperdimensional,
  title={Hyperdimensional computing: An introduction to computing in distributed representation with high-dimensional random vectors},
  author={Kanerva, Pentti},
  journal={Cognitive Computation},
  volume={1},
  pages={139--159},
  year={2009},
}

@Article{SmolenskyTensor1990,
	author = {P. Smolensky},
	title = {Tensor Product Variable Binding and the Representation of Symbolic Structures in Connectionist Systems},
	journal = {Artificial Intelligence},
	year = "1990",
	volume = "46",
	number = "",
	pages = "159--216"
}

@article{smolensky2022neurocompositional,
  title={Neurocompositional computing: From the Central Paradox of Cognition to a new generation of AI systems},
  author={Smolensky, Paul and McCoy, Richard and Fernandez, Roland and Goldrick, Matthew and Gao, Jianfeng},
  journal= "{AI Magazine}",
  volume={43},
  number={3},
  pages={308--322},
  year={2022}
}

@inproceedings{MalsburgAssemblies1986,
  title={Am {I} thinking assemblies?},
  author={C. von der Malsburg},
  booktitle="{Brain Theory}",
  pages={161--176},
  year={1986},
}

@inproceedings{goldreich1999chinese,
  title={Chinese remaindering with errors},
  author={Goldreich, Oded and Ron, Dana and Sudan, Madhu},
  booktitle={Annual ACM symposium on Theory of Computing (STOC)},
  pages={225--234},
  year={1999}
}

@Article{FodorCritical1988,
	author = "J. A. Fodor and Z. W. Pylyshyn",
	title = {Connectionism and Cognitive Architecture: A Critical analysis},
	journal = {Cognition},
	year = "1988",
	volume = "28",
	number = "1-2",
	pages = "3--71"
}

@article{FeldmanConnectionist1982,
  title={Connectionist models and their properties},
  author={J. A. Feldman and D. H. Ballard},
  journal= "{Cognitive Science}",
  volume={6},
  number={3},
  pages={205--254},
  year={1982},
}

@InProceedings{mel1989sigma,
  title={Sigma-{Pi} learning: On radial basis functions and cortical associative learning},
  author={B. W. Mel and C. Koch},
  booktitle = "{Advances in Neural Information Processing Systems (NeurIPS)}",
  pages={474--481},
  year={1989}
}

@article{teyler1986hippocampal,
  title={The hippocampal memory indexing theory},
  author={Teyler, Timothy J and DiScenna, Pascal},
  journal={Behavioral Neuroscience},
  volume={100},
  number={2},
  pages={147--154},
  year={1986},
}

@article{frady2019robust,
  title={Robust computation with rhythmic spike patterns},
  author={Frady, E Paxon and Sommer, Friedrich T},
  journal={Proceedings of the National Academy of Sciences},
  volume={116},
  number={36},
  pages={18050--18059},
  year={2019},
}

@inproceedings{plate1992holographic,
  title={Holographic recurrent networks},
  author={Plate, Tony A},
  booktitle = "{Advances in Neural Information Processing Systems (NeurIPS)}",
  volume={5},
  year={1992}
}

@article{goyal2022inductive,
  title={Inductive biases for deep learning of higher-level cognition},
  author={Goyal, Anirudh and Bengio, Yoshua},
  journal={Proceedings of the Royal Society A},
  volume={478},
  number={2266},
  year={2022},
}

@inproceedings{laiho2015high,
  title={High-dimensional computing with sparse vectors},
  author={Laiho, Mika and Poikonen, Jussi H and Kanerva, Pentti and Lehtonen, Eero},
  booktitle={2015 IEEE Biomedical Circuits and Systems Conference (BioCAS)},
  pages={1--4},
  year={2015},
  organization={IEEE}
}
\newpage
\begin{appendix}
\begin{center}
\textbf{\Large Supplemental material}
\end{center}
\section{Mathematical derivations}
\setcounter{equation}{0}
\setcounter{figure}{0}
\setcounter{table}{0}
\setcounter{page}{1}
\makeatletter
\renewcommand{\theequation}{S\arabic{equation}}
\renewcommand{\thefigure}{S\arabic{figure}}

\subsection{Similarity-preserving properties of embeddings}
\label{sec:residue_kernel}
In the following section, we examine the similarity-preserving properties of our coding scheme. Recall from Section~\ref{sec:space_representation} that our crucial desiderata are that: (1) distinct residue values are represented using vectors which are nearly orthogonal, and that (2) the inner-product between representations of sub-integer values are reflective of a reasonable notion of similarity between the encoded values. There is a robust literature on this topic both within the Vector Symbolic Architectures community \cite{plate2003holographic,thomas2021theoretical,frady2022computing,clarkson2023capacity}, and the broader ML community \cite{rahimi2007random} who often study these techniques under the name ``random features.'' The methods pursued here are in this tradition.

To briefly recapitulate the construction of Equation~\ref{eq:fpe}: fix some positive integer $m$, and let $P(k)$ denote the uniform distribution over $\{0,...,m-1\}$. Define an embedding $g : \mathbb{R} \to \C^{D}$ using the following procedure: draw $k_{1},...,k_{D}$ independently from $P(k)$, and set:
\[
    g(a)_{j} = \exp\left(i \omega k_{j} \right)^{a} / \sqrt{D},\, j = 1,...,D,
\]
where $\omega = 2\pi/m$, and $i = \sqrt{-1}$. To simplify analysis, we here assume that $m$ is odd, in which case the above is equivalent to shifting the support of $P(k)$ to $\{-(m-1)/2,...,(m-1)/2\}$, and defining the embedding $g : \R \to \C^{D}$ component-wise via:
\[
    g(a)_{j} = \exp\left( i \omega k_{j} a \right) / \sqrt{D},\, j = 1,...,D.
\]
The case that $m$ is even is slightly different, but can be handled using similar techniques and the discrepancy does not affect any of our modeling goals.

Our basic claim is that in expectation with respect to randomness in the draw of $k_{1},...,k_{D}$, inner-products between the embeddings of two numbers $a,a'$ recover the periodic sinc-function \cite{smith2011spectral} of their difference. That is:
\[
    \E[\mathbf{g}(a)^{\top}\mathbf{g}(a')^{*}] = \frac{\sin(\pi (a - a'))}{m\sin(\pi(a - a')/m)} := \text{psinc}(a - a'),
\]
This accomplishes goal (1) because, for $t$ an integer which is not an integer multiple of $m$, $\text{psinc(t)} = 0$. Therefore, distinct integers are represented using vectors which are, in expectation, orthogonal. It also accomplishes goal (2), because $\text{psinc(t)} \approx 1$ for $0 < |t| \ll 1$.  The following theorem demonstrates this property more formally, and provides an approximation guarantee for a \emph{specific} instantiation of $k_{1},...,k_{D}$.
\begin{theorem}
    \label{thm:asinc-kernel}
    Fix any $D > 0$ and $\delta \in (0,1)$. For any pair $a,a' \in \R$ such that $a - a'$ is not an integer multiple of $m$, with probability at least $1-\delta$ over randomness in the draw of $k_{1},...,k_{D}$:
    \[
        \left|\mathbf{g}(a)^{\top}\mathbf{g}(a')^{*} - \frac{\sin(\pi (a-a))}{m\sin(\pi(a-a')/m)} \right| \leq \sqrt{\frac{2}{D}\ln\frac{2}{\delta}}.
    \]
\end{theorem}
\begin{proof}
Fix any pair $a,a' \in \R$, and denote for concision $t = a - a'$. Taking an expectation with respect to randomness in $k_{1},...,k_{D}$ and using a well-known calculation from the signal processing literature~\cite{smith2011spectral}:
\begin{align*}
    \E_{k_{1},...,k_{d}}\left[\mathbf{g}(a)^{\top}\mathbf{g}(a')^{*}\right] &= D\E_{k_{1}}[g(a)_{1}g(a')_{1}^{*}] \\
                                       &= \frac{1}{m}\sum_{k_{1}=-\frac{m-1}{2}}^{\frac{m-1}{2}} \exp\left( i\omega k_{1} (a - a') \right) \\
                                       &= \frac{1}{m}\left(\frac{\exp\left(-\frac{i\omega t(m - 1)}{2}\right) - \exp\left(\frac{i\omega t (m + 1)}{2}\right)}{1 - \exp(i\omega t)} \right) \\
                                       &= \frac{\exp(i\omega t / 2)}{m\exp(i\omega t / 2)}\left(\frac{\exp(-\pi i t) - \exp(\pi i t)}{\exp(-\pi i t/m) - \exp(\pi i t/m) } \right) \\
                                       &= \frac{\sin(-\pi t)}{m\sin(-\pi t/m)} \\
                                       &= \frac{\sin(\pi ( a - a'))}{m\sin(\pi(a - a')/m)},
\end{align*}
The third equality follows from the second by noting that the latter is a sum of a geometric series with common ratio $r = \exp(\omega t)$. The fifth line follows from the fourth by recalling the identity $\sin(x) = (e^{ix} - e^{-ix})/ 2i$. In the limit of $t \to 0$, the expression evaluates to $1$, consistent with the normalized inner product of a vector with itself.

To show concentration around this value, consider:
\[
    \mathbf{g}(a)^{\top}\mathbf{g}(a')^{*} = \frac{1}{D}\sum_{j=1}^{D}\exp(i\omega k_{j} (a - a')),
\]
and note that since the complex part of the sum vanishes in expectation, we may consider, without loss of generality, the average of the real-valued quantities: $\left( \cos(\omega k_{j}(a - a')) \right)_{j=1}^{D}$, which are bounded in the range $\pm 1$. Therefore, by Hoeffding's inequality:
\[
    \pr\left(\left|\mathbf{g}(a)^{\top}\mathbf{g}(a')^{*} - \E[\mathbf{g}(a)^{\top}\mathbf{g}(a')^{*}]\right| \geq \epsilon \right) \leq 2\exp\left(-\frac{D\epsilon^{2}}{2}\right),
\]
whereupon we conclude that, with probability at least $1-\delta$ over randomness in the draw of $k_{1},...,k_{D}$:
\[
    \epsilon \leq \sqrt{\frac{2}{D}\ln\frac{2}{\delta}},
\]
as claimed.
\end{proof}

This result can be readily extended to the binding of multiple residue number values. Let $\mathbf{g}(a) = \bigodot_{i=1}^K \mathbf{g}_i (a)$, where each $\mathbf{g}_i(a)$ is instantiated independently. Then, by independence, we observe that:
\begin{align*}
    \E\left[\mathbf{g}(a)^{\top}\mathbf{g}(a')^{*}\right] &= \E\left[\prod_{i=1}^{K} \mathbf{g}_i(a)^{\top}\mathbf{g}_i(a')^{*}\right] \\
    &= \prod_{i=1}^{K} \E\left[ \mathbf{g}_i(a)^{\top}\mathbf{g}_i(a')^{*}\right]
\end{align*}

The implication is that $\E[\mathbf{g}(a)^{\top}\mathbf{g}(a')^{*}] = 1$ if and only if all residue values agree, and zero otherwise. To show concentration around this value, we can again use Hoeffding's inequality, which recovers the same bound on the sufficient dimension.

\subsection{Information-theoretic estimate of required pattern dimension}
\label{sec:scaling_infotheory}

In this section, we describe an information-theoretic estimate on the dimension $D$ necessary to retrieve $n$ patterns within $K$ modules. The main result we aim to show is that $D = \mathcal{O}(n^{(K-1)/K})$; equivalently, the scaling of $n$ for a given $D$ is $\mathcal{O}(D^{K/(K-1)})$. This scaling roughly predicts our empirical results of finding the dimension required to achieve high accuracy, suggesting that the attractor network described here performs close to the theoretical bound.

The minimal total amount of information a network needs to store for denoising an RNS representation with coding range $M$ is $\mathcal{O}(M \ \text{log}(M))$. This results from the requirement of content addressability, i.e., for serving as a unique pointer to one of $n$ patterns, each pattern must at least carry information of the order of $\mathcal{O}(\text{log}(M))$.
For simplicity, we now assume that each module is of size $\mathcal{O}(M^{1/K})$. The total capacity of the network is bounded by the number of synapses, which is $\mathcal{O}(D*K*M^{1/K}) = \mathcal{O}(D*M^{1/K})$ (assuming $K$ is constant), times the capacity per synapse. Under the conservative assumption that the capacity per synapse is $\mathcal{O}(1)$, the dimension is of order $\mathcal{O}(e^{\frac{K-1}{K}\log{(M)} + \log{(\log{(M)})}})$. Thus, the leading order of how $D$ depends on $n$ is $\mathcal{O}(M^{(K-1)/K})$. If the capacity per synapse is assumed to be larger, $O(\log{(M)})$ bits, only the non-leading term cancels and the resulting order of $D$ is still the same.

\subsection{Construction of triangular frames}
\label{sec:hex_frame}

In order to convert a $2D$ coordinate $\mathbf{x}$ into a $3D$ frame $\mathbf{y}$, we first multiply it by a matrix, $\Psi$ whose rows are the elements of a $3D$ equiangular frame: 
\begin{equation}
    \mathbf{y} = \begin{bmatrix}
    -1/\sqrt{3} & -1/3 \\ 
    1/\sqrt{3}&-1/3  \\ 
     0&2/3 
    \end{bmatrix} \mathbf{x}
\end{equation}

(This particular frame is commonly referred to as a `Mercedes Benz' frame due to its resemblance to the iconic symbol.)  A consequence of working with an overcomplete frame is that there may exist multiple values of $\mathbf{y}$ that correspond to the same $\mathbf{x}$. For this frame, the null space of $\Psi^+$ is the subspace spanned by ${[1,1,1]^\intercal}$ -- grounding the intuition that equal movement in all equiangular directions ``cancels out.'' It therefore might seem that triangular frames require extra operations to determine if two coordinates are equal, but here we show how to avoid this consequence.

The core strategy is to choose seed vectors $\mathbf{g}_{i,1},\mathbf{g}_{i,2},\mathbf{g}_{i,3}$ for each modulus $m_i$ that implement this self-cancellation. For a modulus $m_i$, we draw the phasors of seed vectors from the $m$-th roots of unity. However, we further require that, for each vector component, the three selected phases sum to $0 \ (\text{mod} \ 2\pi)$. We then form a hexagonal coordinate vector by binding the three seed vectors:
\begin{equation}
    \mathbf{g}_i = \mathbf{g}_{i,1} \odot \mathbf{g}_{i,2} \odot \mathbf{g}_{i,3}
\end{equation}
By enforcing that the phases sum to $0 \ (\text{mod} \ 2\pi)$, we ensure that positions that have an equivalent $\mathbf{x}$ coordinate are mapped to the same $\mathbf{g}_i$. Observe that Hadamard product binding of phasors is equivalent to summing their phases, and that binding $e^{0i}$ corresponds to adding nothing. Hence, a pair of three-dimensional coordinates whose differences are a multiple of $[1,1,1]$ will be mapped to equivalent vector representations. Finally, we then form the residue number representation for different moduli by binding, as in Eq.~\ref{eq:rnsvecrep}. The presence of multiple modules and self-cancellation properties complement prior work on the efficiency of hexagonal kernels for spatial navigation tasks~\cite{KomerNavigation2020,komer2020biologically}.

The equivalence of certain 3D coordinates also helps us count the number of states. Clearly, the redundancy means that we have less than $m^3$ states, but it also shows us that every position in the hexagonal grid can be represented by a 3D coordinate which contains at least one coordinate equivalent to $0$. There is one state where all coordinates are $0$, $3(m-1)$ states where exactly two coordinates are 0, and $3(m-1)^2$ states where exactly one coordinate is zero. Thus, there are $3m^2 - 3m + 1$ states for the hexagonal lattice, compared to the $m^2$ states for the square lattice.

In the case of square lattices in $2D$, all states occupy an equal proportion of space; however, this is not the case for the hexagonal lattice (see Figure~\ref{fig:hexagonal}A). This is because states with more zero-valued coordinates occur slightly more frequently. To estimate the effect of unequal proportions on the entropy, we directly calculate the Shannon entropy of hexagonal lattices for finite size spatial grids of increasing radius $l$, as an approximation to the infinite lattice. 
We find that even for $l=1000, m > 7$ the hexagonal code has $99$ percent of the entropy of a system that divided all possibilities equally, and that this gap decreases as $m$ grows larger. Thus asymptotically, as $m \to \infty$, the ratio of entropy for hexagonal vs. square grids tends towards $\text{log}_2(3)$. 

\section{Experimental details}
\label{sec:exp_details}
All experiments were implemented in Python involving standard packages for scientific computing (including NumPy, SciPy, Matplotlib). We describe here the parameters and training setup of our experiments in further detail.

\subsection{Scaling in dimension}

For each number of moduli, $K$, we seek to find the smallest dimension $D$ for which our attractor model factorizes its input, $\textbf{p}$, into the correct grid states in a fixed time ($50$ iterations) with high probability (at least $99$ percent empirically). In instances where the network states remain similar over time (at least $0.95$ cosine similarity), we consider that it converged to a fixed point. If such convergence did not occur, we evaluate the accuracy at the last time step.

To evaluate scaling, we first choose our base moduli to be a set of $K$ consecutive primes. We randomly select one of $M$ random numbers to serve as the input and set the grid states to be random. We then evaluate a candidate dimension on the factorization task for a set number of trials ($200$) and check accuracy. We compare accuracy by considering whether the amplitude of the complex-valued inner products are highest for the true factor. If the accuracy is above our threshold, we then evaluate performance of a slightly higher dimension (dimensions evaluated are spaced apart on a logarithmic scale). Once a sufficiently high dimension achieves the accuracy threshold, we assume that the scaling is non-decreasing and use the last successful dimension as the first try.

Finally, we fit linear regression to all data points on a log-log scale to estimate the scaling between dimension and problem size. We report the slopes to estimate the scaling coefficients.

\subsection{Error correction}
\label{sec:error_correction_methods}
\textit{General experimental setup.} We fix in advance the vector dimension, noise level (determined by $1/\kappa$), and number of moduli. Given these parameters, we estimate the empirical accuracy of factorization on an arbitrary input known to correspond to one of the patterns. We use the same method for checking convergence as above, though we increase the maximum number of iterations to $100$. For all experiments in this section, we average over $1{,}000$ trials.

In the case of input noise, the vector $\mathbf{p}$ is multiplied by a noise vector. In the case of update noise, after every time step, each module of the attractor network is corrupted by a von Mises noise update. In the case of codebook noise, all codebooks are corrupted before the start of any iterations.

\textit{Decoding values between integers.} In order to test the ability of the modular attractor network to decode at sub-integer resolution, we fix a spatial resolution $\Delta x$ to decode from. In our experiments, we test $\Delta x = \{1/3, 1/7, 1/15, 1/31\}$, and we also report $\Delta x = 1$ (integer decoding) as a control. Then, using as input a random integer and random multiple of $\Delta x$, we let the modules of the attractor network settle until convergence (as in other experiments). To evaluate accuracy, we test if the resulting output of the attractor network, $\odot_i \mathbf{\hat{g}}_{i=1}^{K}(t)$, is closer to the ground truth RNS representation than to any other value. We test this with a ``coarse-to-fine'' approach: first checking if it is within an integer, and then checking all fractional values within one of that integer. We regard the output as correct if both the integer and fraction match, and incorrect otherwise.

\textit{Estimation of information content from a vector.} To measure the total resolution of our coding scheme in bits, we factor in both the number of states distinguished ($\tau = \frac{M}{\Delta x}$ and the empirical accuracy ($\rho$). To quantitatively estimate this, we report the information decoded in bits according to the following equation~\cite{frady2018theory,kleyko2023efficient}:
\begin{equation}
\label{eq:MI:item}
\begin{split}
I(\tau,\rho) = & a \log_2(\tau \rho ) +  (1-\rho) \log_2 \left( \frac{\tau}{\tau-1} (1-\rho) \right).
\end{split}
\end{equation}
\noindent
A consequence of this equation is that the information decoded is $0$ when the empirical accuracy is at chance ($1/\tau$).

\subsection{Path integration}\label{sec:appendix_path_integration}

\textit{General experimental setup.}
We generate paths using a statistical model simulating rodent two-dimensional trajectories in a 50 $\text{cm}^2$ closed square environment~\cite{raudies2012modeling,banino2018vector}, with $\Delta t = 100$ ms.  The path integration method starts from the ground truth first position $(x_0,y_0)$ which is converted to hexagonal coordinates $(a_0, b_0, c_0)$ (see Section~\ref{sec:hex_frame}) and encoded as an RNS representation $\mathbf{p}(0)$ of dimension $D=3{,}000$ following the method in Section~\ref{sec:space_representation}, for moduli $\{3,5,7\}$. We then factorize $\mathbf{p}(0)$ into $\{\mathbf{\hat{g}_i}(0)\}_{i=1}^K$ to produce the estimated representation $\mathbf{\hat{p}}(0)=\bigodot_{i=1}^K \mathbf{\hat{g}_i}(0)$.

At each time step $t\geq 0$, we aim at estimating the position $(x_{t+1},y_{t+1})$. We give the modular attractor network as input the previous position vector estimate $\mathbf{\hat{p}}(t)$. It is factorized into the residue components $\{\mathbf{\hat{g}_j}(t)\}_{j=1}^K$ that are then shifted according to the velocity $(da_t,db_t,dc_t)$ between $(a_{t}, b_{t}, c_{t})$ and $(a_{t+1}, b_{t+1}, c_{t+1})$. Namely, for each residue module, we build a velocity vector $\mathbf{q}_{j}(t)=\mathbf{g}_{j,1}(da(t)) \odot \mathbf{g}_{j,2}(db(t)) \odot \mathbf{g}_{j,3}(dc(t))$ that is binded to each residue component $\mathbf{\hat{g}_j}(t)$. The estimated position vector is then the binding of the shifted estimated residue components:  $\mathbf{\hat{p}}(t+1)=\bigodot_{j=1}^K \mathbf{\hat{g}_j}(t) \odot \mathbf{q}_{j}(t)$. The estimated position $(\hat{x}_{t+1},\hat{y}_{t+1})$ is chosen to be the position $(x,y)$ in a grid of $50 \times 50$ positions mapping the entire environment, corresponding to the highest similarity between $\mathbf{p}(x,y)$ and $\mathbf{\hat{p}}(t+1)$.

We show the robustness of the path integration dynamics to two different sources of noise. In the case of extrinsic noise (Figure~\ref{fig:path_integration}D), the hexagonal velocity is corrupted by additive Gaussian noise of variance $0.12$. In the case of intrinsic noise (Figures~\ref{fig:path_integration}A and B), the position vector $\tilde{p}_{t}$ is corrupted by binding with a vector sampled from a von Mises distribution with concentration parameter $\kappa=2$.

\textit{Response field visualization.} Given a moduli $m_i$ and a vector $\mathbf{g}_i$, we visualize its response field by computing the similarity of the modular attractor output $\mathbf{\hat{g}}_i(t)$ and $\mathbf{g}_i$ along a trajectory. The periodicity in the distribution of random weights and the hexagonal coordinates produce periodic hexagonal receptive fields whose scale depends on $m_i$. The receptive fields of a given moduli are translations of one another, because the inner product between vector states induces a translation-invariant kernel.

\textit{Connection to sensory cues.} Sensory cues are random binary vectors of size $N_s=D$ that are associated with positions along the trajectory. When the true trajectory reaches a sensory cue, the hippocampal state $\mathbf{\hat{p}_t}$ is updated using the heteroassociation method described in Appendix~\ref{sec:heteroassociation_methods}

\subsection{Heteroassociation}
\label{sec:heteroassociation_methods}
\textit{General experimental setup.} We evaluate our model's performance for pattern denoising using a heteroassociative learning rule~\cite{sharma2022content,chandra2023high}. We consider random binary patterns of size $N_s={D}$. We corrupt the patterns by randomly flipping bits with probability $p_{\mathrm{flip}} \in [0,0.5]$ and associate them to place cell representations using heteroassociation with a pseudo-inverse learning rule. Let $\mathbf{S} \in \R^{N_s \times M}$ be the matrix of $M$ patterns to hook to the scaffold and $\mathbf{H} \in \C^{M \times D}$ the matrix of $M$ position vectors on which to hook the patterns. We associate pattern $\mathbf{s}$ to a place cell representation $\mathbf{p}=\mathbf{H}\mathbf{S}^{+}\mathbf{s}$, where $\mathbf{S}^{+}$ is the pseudo-inverse of $\mathbf{S}$. The model returns a denoised place cell representation $\mathbf{\hat{p}}$ from which we can estimate a denoised pattern by inverting the heteroassociation projection $\mathbf{\hat{s}}=\mathrm{sgn}\left(\mathbf{S}\mathbf{H}^{+}\mathbf{\hat{p}}\right)$.

\textit{Scaling to dimensionality.} We evaluate the impact the dimension $D$ has on the denoising performance in Figure~\ref{fig:heteroassociation}, for a number of stored patterns $M=60 \ (\text{in this case, } 3\times4\times5$) and $210 \ (\text{in this case, } 5\times6\times7)$. For each dimension $D \in \{256, 512, 1024, 2048\}$, we show the evolution of accuracy for different levels of corruption. For a given dimension $D$ and noise level $p_{\mathrm{flip}}$, we denoise a pattern and consider that the denoising is correct if the denoised pattern is closest to the ground truth pattern (in terms of cosine similarity). We repeat over 500 trials and report the accuracy as well as the average similarity (normalized inner product) between the denoised pattern and its noiseless version.

\textit{Superposition of patterns.} We show that our model can denoise a superposition of $n_p$ patterns one at a time, for $n_p \in \{1,2,3,4,5,10\}$. We fix the dimension $D$ to $2{,}000$ and for different values of bit flip probability $p_{\text{flip}} \in [0,...,0.5]$, we run the model on a superposition $\mathbf{s}$ of random binary patterns $\{\mathbf{s}_1,...,\mathbf{s}_{n_p}\}$ of size $N_s=2{,}000$: $\mathbf{s}=\mathbf{s}_1+....+\mathbf{s}_{n_p}$. We run the model $n_p$ times and between each run the denoised pattern is explained away from the superposition \cite{kymn2024compositional}. Namely, for run $r \in \{1,...,n_p-1\}$ we denote $\mathbf{\hat{s}}(r)$ the denoised pattern. The input to run $r+1$ is then $\mathbf{s}(r+1)=\mathbf{s}(r)-\mathbf{\hat{s}}(r)$. We find that the more patterns are superposed, the lower the overall denoising accuracy is. This is due to the fact that when a pattern is incorrectly denoised, explaining away adds noise or spurious patterns to the representation of the superposition which makes the following denoising steps more difficult.

\begin{figure}[h]
    \centering
    \includegraphics[width=0.4\linewidth]{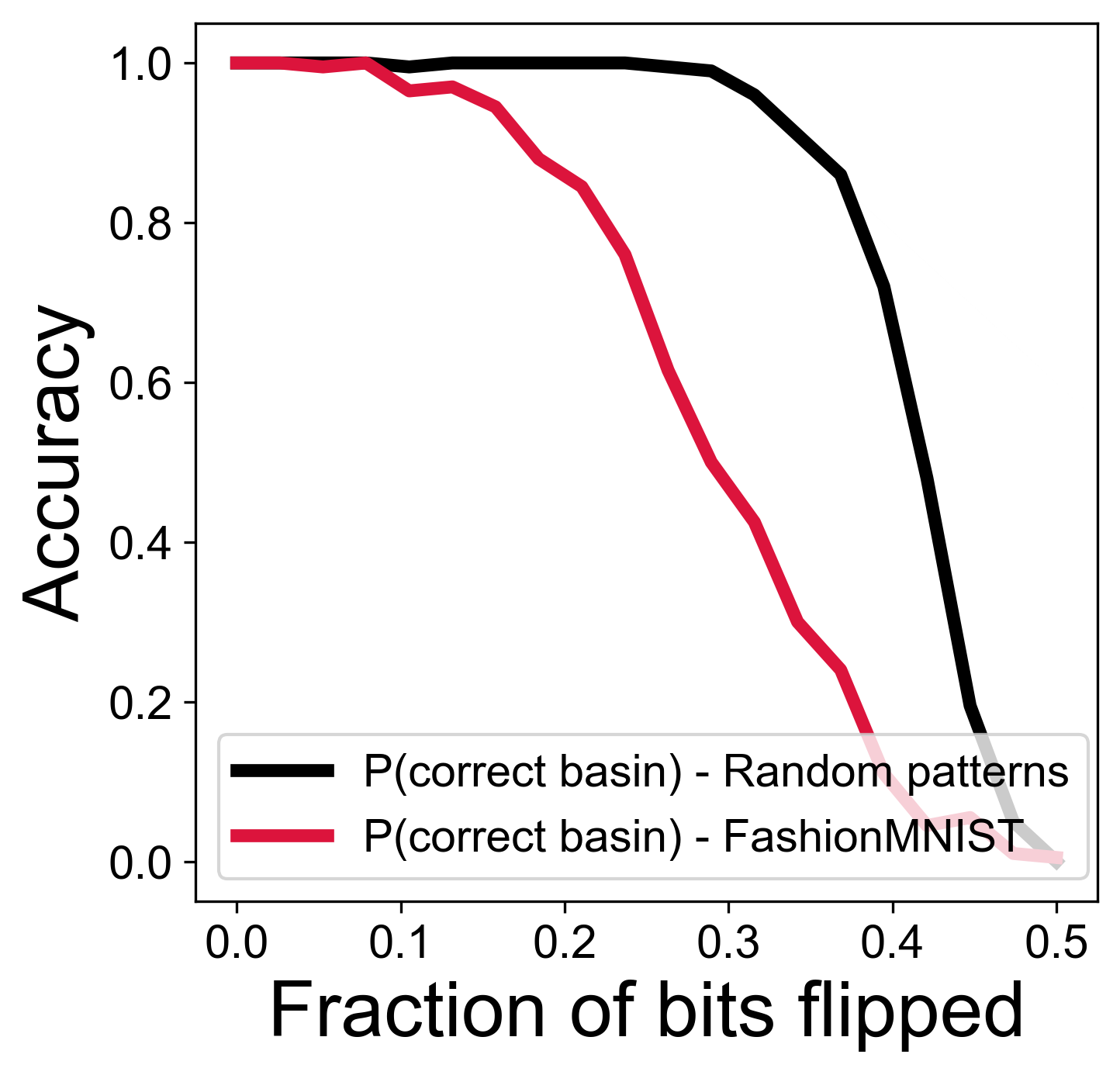}
    \caption{Comparison of heteroassociation on random patterns vs. binarized version of the FashionMNIST dataset. For different levels of corruption, we denoise flattened binarized FashionMNIST images as well and random binary vectors of the same size. The overall denoising accuracy is lower for FashionMNIST. }
    \label{fig:fashionmnistcomp}
\end{figure}

\textit{Comparison to structured patterns.} We evaluate our model's ability to denoise structured patterns. We consider the FashionMNIST dataset, from which we select $105$ images of size $28\times28$ that we binarize by setting pixel values to be $-1$ if below $127$, and $1$ elsewhere. We compare the denoising performance to the performance on random binary patterns of size $28\times28=784$ for fair comparison (Figure~\ref{fig:fashionmnistcomp}).

\section{Additional results}

\subsection{Further visualizations of grid cell modules}

\begin{figure}[h]
    \centering
    \includegraphics[width=0.6\linewidth]{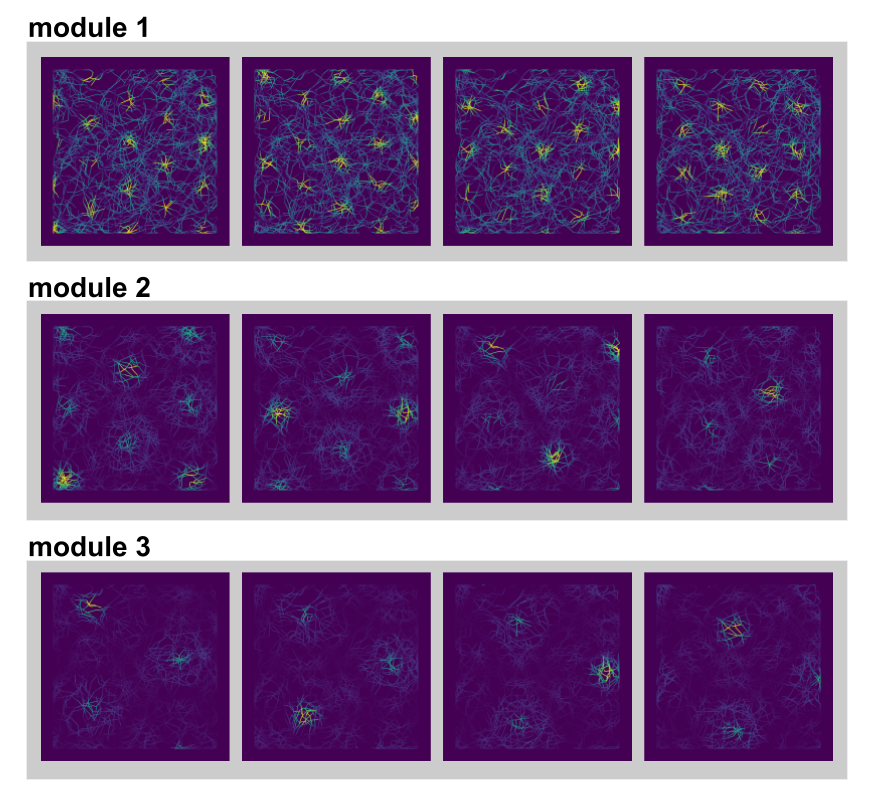}
    \caption{Response fields visualization of $4$ different $\mathbf{g}_i$ in $3$ different modules $m_i=3,5$ and $7$. For a fixed module, the response fields are translations of one another.}
    \label{fig:ReceptiveFields}
\end{figure}

We further visualize the receptive fields for path integration by showing receptive fields from different units taken from the same grid module. We simulate a trajectory that traverses the entire environment and represent the activation of different position vectors along the trajectory. For each modulus $m_i \in \{3,5,7\}$, we show the similarity between $4$ different vectors $\mathbf{g}_i$ from module $m_i$ and the position vectors along the trajectory.  We show in Figure~(\ref{fig:ReceptiveFields}) that the different receptive fields of a given module are translations of one another. 

\subsection{Remapping contexts}

\begin{figure}[h]
    \centering
    \includegraphics[width=0.6\linewidth]{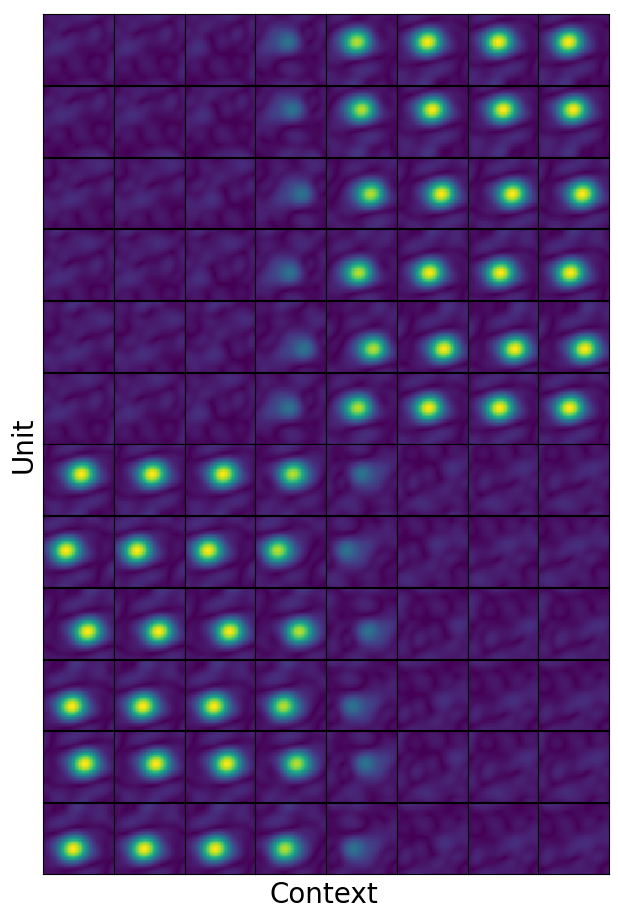}
    \caption{Remapping of place cells depending on context, similar to what was observed experimentally in an experimental study of attractor network dynamics in hippocampus~\cite{wills2005attractor}. }
    \label{fig:remapping}
\end{figure}

We demonstrate that the context vector can serve as a model of \textit{global remapping} in hippocampal place fields, which occurs when there is no relationship between the firing of place cells in different environments \cite{latuske2018hippocampal}. The simplest instance of this is when a place field occurs in context A but not context B, consistent with the observed sparsity of hippocampal activity \cite{thompson1989place}. To model this kind of remapping phenomenon, we consider an instance where there is a gradation of contexts with some phase transition between them; such an instance was observed experimentally \cite{wills2005attractor}. Towards this end, we model linear combinations of these contexts, where the weights each context is given are $\text{sigmoid}(x), 1-\text{sigmoid}(x)$, with x varying from $-5$ to $5$ in $8$ equally spaced increments, and 
with $\text{sigmoid}(x) = 1/(1+\text{exp}(-x))$. To model hippocampal units, we generate units that prefer one of the two contexts and have a random place field location, using its weight vector, or address, as $ \mathbf{c} \bigodot_{i=1}^K \mathbf{g}_i$, and compare its output to that of the context/grid system at each location and context. It is worth noting that the original experiment of \cite{wills2005attractor} also exhibited instances of rate remapping for some units, and so there is certainly additional complexity underlying remapping that is not captured by our simple model.

\begin{figure}[h]
    \centering
    \includegraphics[width=0.7\linewidth]{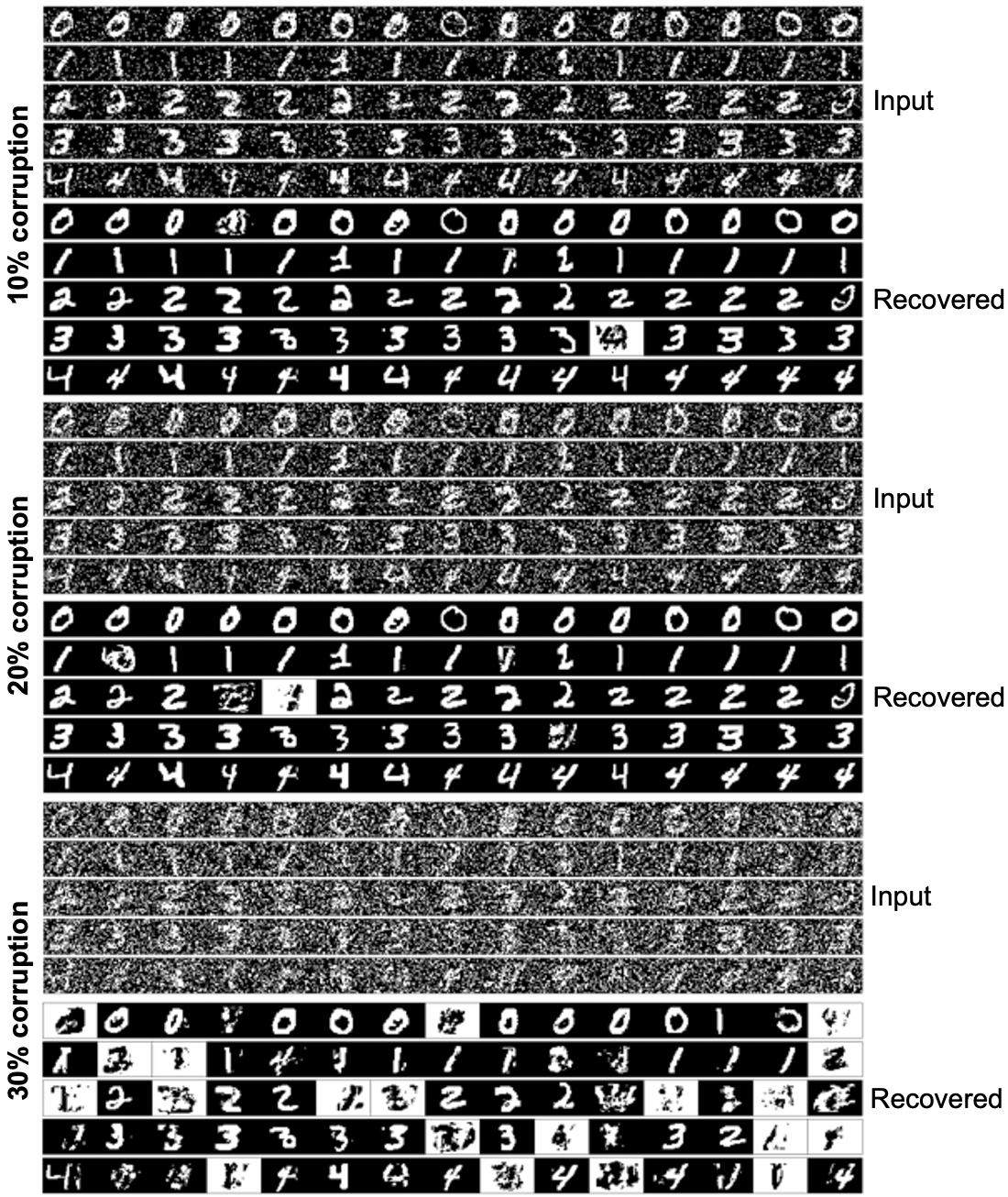}
    \caption{Heteroassociation with contexts on the MNIST Dataset, at varying degrees of corruption. }
    \label{fig:heteroassociation_context}
\end{figure}

\subsection{Storing and retracing sequences}
\label{sec:appendix_sequences}

\begin{figure}[t]
    \centering
    \includegraphics[width=\linewidth]{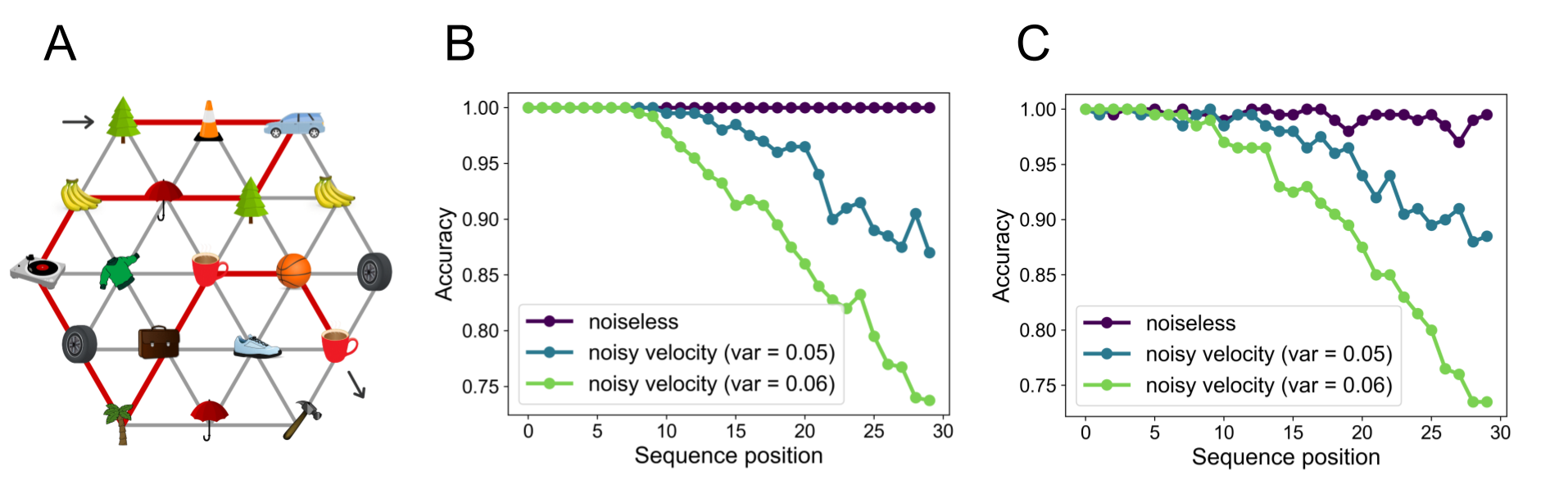}
    \caption{\textbf{Flexible sequence retrieval via path integration in a conceptual space.} \textbf{A)} An example of a hexagonal lattice with sensory observations associated with different states. Having knowledge of the underlying graph enables generalization to new trajectories in the space \cite{behrens2018cognitive,whittington2020tolman}. \textbf{B)} Accuracy of random binary pattern retrieval as a function of position in the sequence for a fixed error rate and one context tag. The noiseless case achieves perfect accuracy, but errors accumulate after incorrect sequence predictions. \textbf{C)} Same as B), but with the additional task of inferring the context tag.}
    \label{fig:sequences}
\end{figure}

We demonstrate that our model can recover sequences by heteroassociation of patterns to positions and path integration in a conceptual space (Figure~\ref{fig:sequences}A). This is consistent with the postulated role of the hippocampal formation in performing navigation in conceptual spaces~\cite{constantinescu2016organizing,bellmund2018navigating}, and the role of entorhinal cortex in generating sequences of neural firing in hippocampus~\cite{schlesiger2015medial,yamamoto2017direct}. 
To evaluate our attractor model's fidelity at sequence memorization and retrieval, we simulate trajectories to form sequences of random binary patterns and recall the sequence using the path integration mechanism following the method in Section~\ref{sec:path_integration}, for $D=10{,}000$ and moduli $\{3,5\}$. We add extrinsic noise to the velocity input, which accumulates along the trajectory and induces a drift. This implies that patterns at the end of sequences are less well recovered than ones at the beginning (Figure~\ref{fig:sequences}B and C).

\end{appendix}

\end{document}